\documentclass[sigplan,10pt,nonacm]{acmart}  
\microtypesetup{protrusion=true,expansion=true,factor=1100,stretch=40,shrink=40,step=1}
\usepackage{booktabs}
\usepackage{array}
\usepackage{xspace}
\usepackage[framemethod=TikZ]{mdframed}

\newcolumntype{L}[1]{>{\raggedright\arraybackslash}p{#1}}
\newcommand{\sys}{Worldline\xspace}
\newif\ifextended
\extendedtrue  
\newif\ifanon \anonfalse   
\newif\ifacmlook \acmlooktrue 

\newmdenv[linewidth=0.6pt,roundcorner=2pt,
          innertopmargin=5pt,innerbottommargin=5pt,
          innerleftmargin=7pt,innerrightmargin=7pt,
          skipabove=6pt,skipbelow=6pt]{principlebox}
\newenvironment{principle}[1][The Exact-State Admissibility Condition]{%
  \begin{principlebox}\footnotesize\textbf{#1.}\par\smallskip}{%
  \end{principlebox}}

\begin{document}

\title[Stateful Worlds, Stateless Elasticity]{Stateful Worlds, Stateless Elasticity:\\ Exact-State Serving for Interactive World Models}

\author{Jin Li}
\authornote{Work done while on a leave of absence from Harvard University.}
\affiliation{\institution{Harvard University}\city{Cambridge}\state{MA}\country{USA}}

\author{Jiawei (Alexon) Chen}
\affiliation{\institution{Independent Researcher}\country{}}

\begin{abstract}
A persistent interactive world model keeps its running state resident
on the GPU that serves it: a multi-gigabyte attention cache, almost all
of it rewritten at every generation step. That state cannot be
recomputed in interactive time or approximated without changing the
world, so a live session pins its device: one user, one GPU. We show the
pin is a scheduling problem. \emph{WorldMove} moves a live session under
one guarantee: the destination is bit-identical to the source, or
nothing is installed. It relocates the cache in $18.8$\,ms same-node,
$101\times$ faster than save/load, to our knowledge the only other deployed exact-class primitive. It
holds a checksum-verified $92.1$--$94.8$\,Gb/s on a $100$\,Gb fabric. At that rate the cache fits inside one interactive block. Migrating an
\emph{actively generating} session, it converges at a block boundary,
and the destination continues the world bit for bit. An admissibility condition
decides each move. The move must complete inside the readout horizon, over
bandwidth that covers the state plus its dirty rate. The condition lifts
to a fleet schedulability test; the consolidation loop it governs
executed $48$ of $48$ fleet migrations bit-identical across two providers. Two constraints are structural. Bit-exactness survives only
inside a controlled configuration of one GPU architecture, so moving the
state is the only way to preserve it exactly in interactive time. On this fabric, verification cannot hide inside
the wire. Receive-path checksums stall the transport at protocol-timer
timescales under fan-in, and, separately, unscheduled incast silently collapses a
receiver while every delivered byte stays correct. An incast-aware
admission controller holds zero misses to $1.4\times$ offered load and
sheds overload as explicit rejects. A lossless GPU codec widens the
admission gate to fabrics that raw motion cannot use. We exercise the serving loop and the mover
separately, each end to end. Their composition on one fabric is
unbuilt. Exact-state elasticity is a joint scheduling problem over
transport and verification.

\end{abstract}

\maketitle

\section{Introduction}
\label{sec:intro}

\looseness=-1
An interactive world model's running state lives on the GPU that
serves it: an attention cache of $1.67$\,GB on one production engine
(\S\ref{sec:measure}), three-quarters of it rewritten every
generation step, and growing with per-session context length. The model
advances that state one
action-conditioned block at a time, and once a user is inside the world the
state \emph{is} the running computation. It cannot be recomputed from a
prompt, re-derived from its log in interactive time, or approximated without
visibly changing the world. The contract a viewer holds is \emph{exact
continuity}: the next frame must be the one this exact trajectory would
have produced. Because no deployed serving primitive can relocate that state under that contract, a live
session \emph{pins} its device for its whole lifetime: one user, one GPU. The
vendors say so plainly. World Labs runs one interactive
stream per dedicated H100 and prices persistence as ``neither feasible nor
economically viable given today's computing infrastructure''~\cite{rtfm};
Decart, operating Oasis in production, calls serving cost ``the hidden
bottleneck to releasing generative video in production''~\cite{oasisserve},
and its successor ships at \$0.02 per simulated second (\$72 per user-hour)~\cite{oasis3price}.
At one device per stream, $10^4$ concurrent sessions provision $10^4$ GPUs, the same cost cliff that forced virtualization into the datacenter two decades
ago.

We measured the pin directly. Under the exact-continuity contract every
lossy handle we evaluated (eviction, quantization, low-rank projection,
cross-session dedup) breaks the world ($60$--$92$\,dB divergence,
\S\ref{sec:measure}), so the destination must receive the whole state
intact, by moving it. Lossless compression, the remaining
contract-preserving handle, shrinks the wire payload but still delivers
the whole state. Measured in \S\ref{sec:eval-xnode}, it widens the
admission gate.
Moving it is an \emph{operator} need before it is a workload need
(defragmentation, maintenance evacuation, failover, heterogeneous
placement), as live VM migration was before the cloud demanded
it~\cite{clark2005migration}. The demand has since arrived. A production
streaming-video fleet live-migrates stateful sessions tens of times per
two-minute window as routine control~\cite{turboserve}.

That fleet moves state without verifying its bytes, safe only while the
fabric is clean. \S\ref{sec:eval-m4} reports one that was not.
Exact-state elasticity therefore puts motion \emph{and its verification}
under one contract, so a fleet can depend on the state it moves.

No existing elasticity abstraction supplies this, because stateless
routing has no vehicle that carries per-session state. VM pre-copy~\cite{clark2005migration}
needs a small writable working set, and this object rewrites $\sim$$3/4$ of
its cache every step; append-only KV migration~\cite{llumnix} hides an
immutable prefix, and this cache is mostly mutable.
Checkpoint/restore~\cite{criu} trades away interactive continuity.
Lossy reduction is the family of handles measured to break the world,
the \emph{kill-family} of \S\ref{sec:measure} (each neighbor,
\S\ref{sec:related}). The two escape hatches that
spared every prior continuous-media system are closed here. An
action-conditioned world cannot \emph{buffer} what depends on an action not
yet taken, and \emph{native-rate replay} breaks across GPU architectures
at the hardware level (the same seed and actions diverge at the first
block even with the software stack matched to the cuDNN patch version,
\S\ref{sec:eval}). Software-emulated re-execution can be made bit-exact
offline~\cite{bitexactverif}, but cannot serve an interactive readout
horizon.
On a mixed fleet, the only operation that preserves exactness in interactive time is to move the
bytes, and \S\ref{sec:eval-xnode} measures that motion, verified, within $6\%$ of
the mover's own verify-off ceiling.

Not every generated world needs this. A world exported as an explicit asset
(Gaussian splats and meshes~\cite{marble}) has canonical bytes. The regime
that needs exact-state elasticity is the frame-model class (live
activations, owned by one session~\cite{rtfm}; the window-rewrite engines
measured here rewrite that state every step,
\S\ref{sec:measure-dirty}). As per-session context length outgrows HBM's $\sim$$26\%$/year
growth~\cite{hbm}, the pin tightens. Table~\ref{tab:necessity} (\S\ref{app:boundary}) names
each present workload's escape and the force that closes it.

\looseness=-1{}The pin is a scheduling problem, and \sys{} supplies the primitive so that a session keeps its worldline while the
hardware underneath changes.
\emph{WorldMove} makes the resident cache a
first-class movable object (flatten, transfer, verify, rehydrate) with
one guarantee: \textbf{the destination is bit-identical to the source, or
nothing is installed}. Same-node it moves the live cache in $18.8$\,ms,
$101\times$ faster than save/load, to our knowledge the only other deployed exact-class primitive
(\S\ref{sec:eval-m4}). The same code path moves a text-LLM KV
cache with a token-for-token-identical continuation
(\S\ref{sec:eval-crossmodel}).

\emph{When reconstructing a
session's resident state from its history costs more than the interaction's
latency horizon, elasticity must \textbf{move} that state instead of
rebuilding it}. That rule governs the design.
Demand paging managed a \emph{reducible} resident
set~\cite{kilburn1962,denning1968}; under exact
continuity nothing may be evicted or approximated, so the fundamental
primitive becomes \textbf{exact-state motion} of a stationary generative
working set, managed under a deadline (\S\ref{sec:law} develops the
inversion, the admissibility condition, and its fleet test).

A serving substrate for these worlds must schedule resident \emph{state}.
The scheduled unit is the session object, and requests arrive against it. \sys{} treats a session as a typed, contract-bearing state
object behind a narrow six-verb API
and drives WorldMove fail-closed (\S\ref{sec:design}, Table~\ref{tab:api}).

This paper makes three contributions. First, we characterize a new runtime resource,
readout-irreducible resident GPU state, to the point where
its motion can be scheduled: the kill-family of lossy handles, the
dirty structure, and the update-rule classifier that decides which
engines share the regime (\S\ref{sec:measure}). Second, we build the discipline
that governs its motion: \sys{}, a working serving substrate with WorldMove, a bit-exact mover, and the
admissibility condition that decides when a move is legal and lifts to a
fleet schedulability test. The consolidation loop it governs executed
$48/48$ bit-exact across two providers, plus a separate move against an
actively generating session (\S\ref{sec:law}, \S\ref{sec:eval}). Third, we find that
verification is a \emph{second scheduling plane}: at datacenter rates
its placement decides fan-in stability, an unscheduled incast collapses
a receiver even though every delivered byte is correct, and an incast-aware
admission controller holds zero misses to $1.4\times$ offered load
and sheds overload as explicit rejects
(\S\ref{sec:eval-fleet}, \S\ref{sec:eval-verif}).

\section{The Workload}
\label{sec:background}
The workload defines the readout horizon $H$
that every later admission decision uses. A causal video world model advances a sliding multi-GB attention-KV window
one action-conditioned \emph{block} at a time (persistence over an hour
of interaction requires attending to contexts of well over $100$M
tokens~\cite{rtfm}). The unit of compute is one
denoising step of one block (a \emph{quantum}; anatomy and costs in
\S\ref{sec:measure}), and decode is a separable cost with its own quality
ladder.

Five tenant classes share this substrate: interactive human sessions
(C1: hard per-block deadlines at frame rate; Oasis served this class to
a million users in its first three days under five-minute session
caps~\cite{oasisusers,oasiswiki}, and Google's Project Genie rations it
to $60$-second sessions~\cite{geniesessions}), planner fan-out (C2a: MPC
or tree search forking short branches off a live state, latency-bound),
rollout cohorts (C2b: episode-parallel RL, wave-synchronous), bulk data
generation (C3: throughput, no deadline), and deterministic replay (C4:
re-execution for debugging, evaluation, and provenance). Their deployment idiom today, and our baseline, is dedicated,
statically provisioned GPUs per class. Odyssey prices the resulting
infrastructure at \$$1$--\$$2$ per user-hour on its H100
clusters~\cite{odysseycost}.
\section{Measurement Study}
\label{sec:measure}

Three facts run through the rest of the paper. First, the serving quantum is
fixed-shape and non-preemptible. Second, every lossy handle we evaluated breaks the world (the
kill-family). Third, generation rewrites most of the cache every step (the
dirty structure).

\subsection{Anatomy of a serving quantum}
\looseness=-1
A distilled causal world model generates video in \emph{blocks} (12
frames; 3 latent frames) through a fixed denoising schedule. On an RTX
5090 running Matrix-Game~2.0, four uniform quanta of 93.9\,ms p50 (three
denoise steps plus a context-KV pass of equal cost) yield 375.6\,ms
blocks, 32\,fps against a 400\,ms release period with
6\% slack at full quality. Per-session state is 5.32\,GB of KV caches,
bounding co-residency at 4--5 sessions per 32\,GB device. State is the
binding constraint before FLOPs. Of that resident footprint, the
\emph{movable} session subset (the bit-exact cache set a migration must
carry) is a fixed $1.67$\,GB across $420$ tensors, constant in session
age. The migration-cost model of \S\ref{sec:eval-m4} transports this
subset.

\looseness=-1{}Two features of that timing reach the scheduler. First, within-run
quantum jitter is tiny, but across-run variance is $2.3\times$ larger, and
near-deterministic quanta \emph{phase-lock} against fixed release grids,
a three-phase latency beat that a random-collision model over-predicts
by $\sim$$2\times$. Second, quantum cost tracks the request's conditioning
window ($85.8$ vs $93.3$\,ms for interactive- vs bulk-shaped requests),
so a capacity model keys on request shape.

\subsection{Quality, decode, and state}
Three further measurements complete the picture. First, quality is a
discrete tier ladder whose cost and floor semantics split by contract
class. Second, decode is another workload that reaches $30$\,fps only
with a distilled small decoder. Finally, the state compresses along the
quality axis but cannot be reduced lossily under the divergence
contract. Co-batching is a
weak knob, and each feasibility row is an \emph{operating point} of
device, placement, and knob settings (Appendix~\ref{app:serving}).
Degradation is a contract axis, declared at
admission with the quality floor.

\subsection{The dirty structure}
\label{sec:measure-dirty}
\looseness=-1
The third fact is the write pattern. An action-conditioned window-rewrite
world advances a full generation step even on a null action and rewrites its
attention window in place. Self-measured on Matrix-Game-2, the per-step writable-working-set (dirty) fraction is $\rho_{\mathrm{WWS}}{=}0.71$--$0.78$ of the cache, and at the measured $0.783$\,steps/s that is a dirty rate $D{=}0.93$\,GB/s. This is a property
of the \emph{update rule}. An append-only serving
loop measures $0.000$ on the same instrument, and a third model family,
the KV-recache engine LongLive-1.3B~\cite{longlive}, measures $0.727$ at
every steady-state block boundary on its $3.545$\,GB cache tree under
the same estimator. The LongLive cache also measures $0.249$ during window fill,
append-like, and $0.727$ once the window is full. A
single cache exhibits both regimes, so the update rule is the
classifier for which results transfer. This paper measures the
window-rewrite class. Engines whose persistent state is written once and
thereafter selected rather than rewritten (frame-bank designs such as
RTFM~\cite{rtfm}, per its public description) should land in
the low-dirty cell with pre-copy open (a prediction, untested here).
Avoiding forced exactness carries its own cost: a worse
silent-corruption profile. The roll geometry below scrubs half of
this cache every block, but a frame bank is never rewritten,
so a silently corrupted stored frame conditions every subsequent
generation and the corruption half-life goes from about one block to
unbounded. In that cell an integrity check on transfer matters even
more. Two consequences follow. Pre-copy convergence is governed by $\rho_{\mathrm{bw}}{=}D/B$
(the condition of \S\ref{sec:law}, measured on both sides in
\S\ref{sec:eval-m4}), and the movable set itself is \emph{stationary},
which makes fleet packing a stationary problem.

The movable set is also an \emph{allocation}: $\ge$$19.4\%$ of it is
written by no step and read by none (unwired
action-conditioning caches), so a manifest-defined motion set could ship
$\ge$$19.4\%$ fewer bytes. This is a projection, and every constant in this
paper is measured on the $1.67$\,GB allocation as shipped
(leaf-level accounting, TR~\S E).

The window geometry also bounds what a corrupted byte can do. Each
block overwrites half the window before any read reaches it, so a flip's
survival depends first on its position within the window and only
secondarily on its bit significance. A
placement-pinned single-bit-flip battery ($n{=}24$ stratified) matches
the geometry's prediction, with $0/12$ flips in the overwritten half ever visible,
erased bit-exactly within one block, while $8/8$ exponent- and
sign-class flips in the read half surface at the next block (design and
the mantissa cells, TR~\S E). Only the read-before-overwrite half needs
integrity protection; the write-once conditioning KV outside the window
has no overwrite path and needs its own guard.

\section{Exact-State Schedulability}
\label{sec:law}

\looseness=-1{}Classical real-time scheduling draws its power from three relaxations: jobs are
preemptible, divisible, and approximation-tolerant. The measurements of
\S\ref{sec:measure} remove all three at once. Relocating an in-flight stateful
generative session is therefore a new \emph{scheduling regime}: under an
exact-continuity contract a migration is scheduled as a non-preemptive,
fixed-size, bit-exact real-time transfer task, and an admissibility condition is what remains.
With nothing to evict, the only question is \emph{where the whole
state is resident and by when}. (Irreducibility is scoped to the bit-exact
tier T0 of \S\ref{sec:design}; weaker contracts restore a reducible
subset, \S\ref{sec:eval}.)

\label{sec:law-regime}
The job has three properties no classical movable-work model carries
together, each of which relaxes exactly one classical assumption
(\S\ref{sec:related}). \emph{Atomic}: nothing is installed until the destination matches a
byte-identical fingerprint, and a paused move drains its horizon without
shipping bytes (in overlap mode the bytes already shipped go stale at
$D$), so a started move is modeled as running to commit or abort. \emph{Fixed-size}: the $1.67$\,GB/$420$-tensor movable
subset (\S\ref{sec:measure}) cannot be shrunk by eviction,
approximation, or dirty-page dropping (the kill-family), so the scheduler
can only place the job. \emph{Bit-exact}: all bytes must cross and
verify before the horizon or the resumed trajectory is undefined
(\S\ref{sec:law-necessity}). Delivering part of the state earns
nothing.

\subsection{The exact-state admissibility condition}
\label{sec:law-condition}
\label{sec:law-sufficiency}\label{sec:law-fleet}

\looseness=-1
The object has three parts: resident state $S$ (here the GPU-resident
attention-KV cache), a generation step that rewrites $S$
deterministically given a fixed architecture, seed, and implementation,
and a readout predicate $R$, true when resuming from the delivered state
is contractually indistinguishable from uninterrupted execution, byte
equality at the bit-exact tier T0 and a bounded divergence under softer
tiers. The horizon $H$ is a contract parameter each class declares
(C1's block release period, $400$\,ms here; C2a's branch budget; C2b's
wave barrier), read as a stall bound for a frozen stop-and-copy and as a
state-age bound for a live overlap move. $T_{\mathrm{migrate}}$ below
folds verification in as a \emph{cost}; \S\ref{sec:eval-verif} shows
it is also a \emph{placement}.

\begin{principle}[The Exact-State Admissibility Condition]
Let a stateful generative model carry resident state of size $S$, mutate it at
dirty-rate $D$ (bytes/s rewritten by ongoing generation), and serve under a
readout contract with horizon $H$ as defined above. Over a
fabric of transfer bandwidth $B_{\mathrm{fabric}}$, let $T_{\mathrm{migrate}}(S)$
be the wall-clock time to extract, transmit, verify, and commit $S$.
A migration is admissible \emph{only if} it completes within the contract horizon
and the fabric sustains the resident state plus its dirtying,
\begin{equation}
\label{eq:law}
T_{\mathrm{migrate}}(S) \;<\; H
\qquad\text{and}\qquad
B_{\mathrm{fabric}} \;\ge\; \frac{S}{H} + D ,
\end{equation}
with the contract's divergence bound established separately by the
continuation predicate $R$ (at the bit-exact tier T0, $R$ is byte
equality).
\end{principle}

\noindent The flux clause is the deadline gate of \eqref{eq:gate} rearranged,
and $D$ is charged in the \emph{live overlap} mode, where generation keeps
rewriting the cache while bytes are in flight. A block-boundary
stop-and-copy is the $D{\to}0$ special case: a frozen source dirties
nothing, and the whole $T_{\mathrm{migrate}}$ is paid as user-visible
stall against the same horizon.

\looseness=-1{} The inequality is elementary. Its force is that for
this object every term is adversarial. Classical migration keeps each one
benign: a small writable set, a reducible payload, an approximate resume.
Exact-state serving makes each one binding: $D$ rewrites most of the cache
every step, so no incremental transfer beats a whole-state move
(\S\ref{sec:measure-dirty}); $S$ cannot be reduced without breaking $R$
(\S\ref{sec:measure}); and $T_{\mathrm{migrate}}$ hides a verification
placement that can destabilize the fabric (\S\ref{sec:eval-verif}). It looks
like Liu and Layland~\cite{liulayland} and behaves like nothing classical
schedulers admit. In either direction there is a hard \emph{deadline
gate}
\begin{equation}
\label{eq:gate}
L^{*} \;=\; \frac{S}{B_{\min}},\qquad B_{\min}=B_{\mathrm{fabric}}-D ,
\end{equation}
the minimum admissible horizon. Contracts with $L<L^{*}$ are
\emph{inadmissible} on that fabric, and a provisionable region sits above
it. The clauses separate two failure modes, a \emph{timing}
bound (deliver before the state goes stale) and a \emph{flux} bound (outrun
the bytes generation keeps dirtying, or never catch the moving target). A
system can satisfy one and violate the other.

\label{sec:law-necessity}
Both clauses are necessary under the bit-exact contract. A commit after $H$
delivers a state already past the contract's staleness bound (with $\sim$$3/4$ of
the cache rewritten per step there is no clean prefix, so the resume is
undefined rather than slow~\cite{vestal}), and the flux clause is iterative
pre-copy convergence~\cite{clark2005migration} met at its adversarial
limit. With no sufficient statistic to ship instead, the move cannot be made
incremental: even a convergent pre-copy moves essentially all of $S$,
and the deadline binds. Appendix~\ref{app:tiers}'s subject is the
weaker T1/T2 tiers, where lossy compression or replay become
legal again. Here we need only
the criterion for when motion is forced.

\paragraph{The motion criterion.}\looseness=-1{} The admissibility condition and the escape
table compose into one test that classifies any served object. State must
\emph{move} (reduction, replay, and restart all fail) exactly
when three escapes close. First, no contract-preserving reduction lets the
destination hold less \emph{information} than the whole state (the
kill-family of lossy handles breaks $R$, \S\ref{sec:measure}; a lossless
codec shrinks the wire payload but still delivers the whole state,
\S\ref{sec:eval-xnode}). Second, reconstruction
exceeds the horizon (replay cost grows with session history while transfer
cost is constant in it, \S\ref{sec:eval-fleet}). Finally, the trajectory itself
carries the value, so a restart is semantic loss. When all three hold, the
move's admissibility is set by the deadline gate of
\eqref{eq:gate}. The update rule is the
classifier for membership (\S\ref{sec:measure-dirty}). Table~\ref{tab:necessity}
(\S\ref{app:boundary}) holds the per-workload audit against real
objects.

\paragraph{Sufficiency.}
When \eqref{eq:law} holds under \emph{WorldMove}'s stated T0
preconditions, \sys{} is the constructive proof.
\emph{WorldMove} meets the condition on real models with an atomic
extract/transmit/verify/commit
(Figure~\ref{fig:arch}) that
flattens the live cache, transfers it (same-node \texttt{cudaMemcpy};
cross-node over a checked socket), CRC-verifies, and rehydrates. It
commits only on a byte-identical fingerprint, so a torn state can never
be installed. The constructive instance is a block-boundary
stop-and-copy of the whole state, which the live pre-copy mode of
\S\ref{sec:eval-m4} converges to as its frozen final round. The model enters
only through the condition's constants. WorldMove moves bytes that a
fingerprint checks, with no knowledge of the model that produced them
(two-family demonstration, \S\ref{sec:eval-crossmodel}).

\paragraph{Fleet schedulability.}
\looseness=-1
A fleet migrates many sessions over a \emph{shared} channel, and a move
in flight holds its channel share to commit or abort, because the
deadline gate makes yielded time unrecoverable. Moves therefore block
and interfere as non-preemptive real-time jobs do, and the condition
lifts to a fleet test, the standard
deadline-monotonic response-time test~\cite{jeffay91} specialized to
atomic state motion (the specialization walk, with its blocking and
interference terms, is Appendix~\ref{app:fleettest}). For a homogeneous
fleet this is \eqref{eq:law} at fleet scale, a derivable rule (we do not
claim the bound is tight):
\begin{equation}
\label{eq:nmax}
  N \;\le\; \frac{H\,B}{C}, \qquad C = S + D\,H .
\end{equation}

Read on the Matrix-Game-2 (MG2) cache of \S\ref{sec:measure} ($C{=}S{+}DH\approx2.04$\,GB), $N_{\max}{\approx}2.45$\footnote{The
fluid bound $HB/C$ drops the fixed per-move setup cost, a measured
$\sim$$57$\,ms intercept in the size-ladder fit, and so overcounts by
one slot at small state (charged with the measured wall it gives $1.62$
here, admitting $N{=}1$); the contended response-time measurement of
\S\ref{sec:eval-fleet}, $T(2){=}415$\,ms${}>400$, overrides the fluid
estimate.}
at $100$\,Gbps, ${\approx}9.8$ at $400$. Below $B{\approx}C/H{\approx}41$\,Gbps
even one migration overruns the deadline, the per-session gate recovered
as the $N{=}1$ corner (full bandwidth dependence, TR~\S E). Beyond the gate,
prewarm provisioning sets attainable oversubscription, and added bandwidth
no longer moves it (\S\ref{sec:law-corollary}). The gate assumes raw byte motion. The
lossless codec of \S\ref{sec:eval-xnode} replaces $S$ with the
compressed payload $rS$ plus codec time, widening the admissible region by
the compression ratio, so the $25$\,Gbps fabric that raw motion cannot use
($553$\,ms) becomes admissible ($\sim$$370$\,ms).

Empirically, the serialized response-time structure
holds at three operating points, including full payload at datacenter
rate, where the linear response places the $H{=}400$\,ms
admission integer at the model's own boundary
(\S\ref{sec:eval-verif}, \S\ref{sec:eval-fleet}). Under live churn the dirty term is measured per-move
(\S\ref{sec:eval-m4}) but not yet inside
a serialized fleet schedule.
This result comes with boundaries. Under a readout contract, the
condition is stated and validated for resident-state generative models,
with bit-exact transfer verified \emph{within} an architecture (T0). The necessity
argument is at systems rigor (argued necessity, a working construction
for sufficiency, a measured instantiation) and is not a formal proof. Companion work develops the abstract machine for $R$ and the converse
capacity bound. The condition fixes a hard, measurable admissibility
boundary. A scheduler without the bandwidth
and horizon budget of \eqref{eq:law} cannot serve the object exactly.
\section{Design and Implementation}
\label{sec:design}

\begin{figure*}[t]
  \centering
  \includegraphics[width=0.82\textwidth]{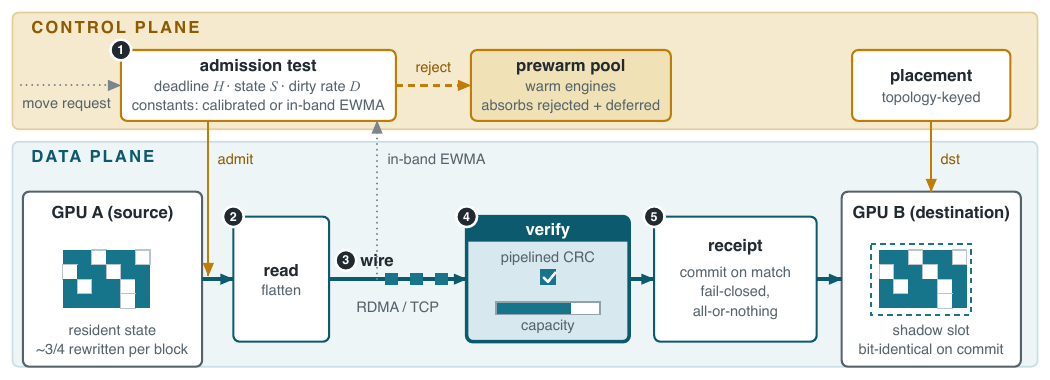}
  \caption{\sys{}: a thin control plane over a state-motion data plane
  whose commits are receipt-checked. One admitted move: admission and placement (1), flatten
  (2), transfer (3), pipelined CRC on a lane with its own capacity (4),
  fail-closed commit on receipt match (5); rejected moves fall to the
  prewarm pool.}
  \label{fig:arch}
\end{figure*}

\looseness=-1{}\sys{}'s control plane is thin
(Figure~\ref{fig:arch}).
Admission is contract-driven. A session declares its
class (C1/C2a/C2b/C3/C4, plus a system maintenance class C5 whose
refresh quanta are scheduled like any other work), a
deadline or period where the class implies one, a quality floor
(minimum denoise steps), a determinism tier (T0 bitwise, T1
schedule-stable, or T2 statistical), and a decode tier. \S\ref{sec:eval}
treats T0's scope and the cross-architecture drop to bounded T1. The type
system rejects incoherent combinations (e.g.\ C4 replay at T2).
\emph{Scheduling} is EDF over quanta with class structure.
Hard-deadline classes own fixed-rate release grids, planner classes get
soft-deadline EDF membership, and bulk fills
residual quanta behind a guard band sized by headroom.
Proactive degradation moves a session down
its contracted quality ladder when its feasibility margin
reaches zero (block-boundary yield). \emph{Admission} checks a candidate's
steady-state quanta against the device's residual at the
session's quality floor. Release grids are warm-anchored (the grid
opens after the measured first-block transient) and held to
queue-depth-1 frame-skip semantics, so lateness cannot cascade.
\emph{State} lives where the session lives. Working sets stay resident
for contract lifetime, forks are in-place KV copies on the parent's
device (\S\ref{sec:eval}), and component-aware
eviction-by-truncation (\S\ref{sec:measure}) is the pressure valve.

\begin{table}[t]\centering\footnotesize
\setlength{\tabcolsep}{4pt}
\caption{The session API. Every verb returns a checkable result (a typed manifest, a verdict, or a boolean), and \texttt{migrate} stages into a shadow slot, so commit or rollback is the only exit.}
\label{tab:api}
\begin{tabular}{@{}l l L{4.0cm}@{}}
\toprule
verb & returns & contract behavior \\
\midrule
\texttt{register\_state} & manifest & binds the live tree to a typed manifest: per-leaf path/dtype/shape/bytes, topology hash, determinism tier \\
\texttt{admit} & verdict & runs the admission arithmetic of \S\ref{sec:law-condition} against the deployment's own profile rows \\
\texttt{migrate} & manifest & drives WorldMove into the shadow slot under the fail-closed fingerprint check; the live handle is untouched \\
\texttt{verify} & bool & re-fingerprints the staged copy against the register-time oracle \\
\texttt{commit} & manifest & re-verifies, then atomically swaps the staged copy live \\
\texttt{rollback} & manifest & atomic undo; the pre-migration tree stays authoritative and the staged copy is dropped \\
\bottomrule
\end{tabular}
\end{table}
\paragraph{The session API.} One
\texttt{WorldlineSession} object drives the whole lifecycle through the six
verbs of Table~\ref{tab:api}. The \texttt{migrate} verb stages into the
shadow slot of Figure~\ref{fig:arch} and never touches the live handle,
so a half-migrated state is unrepresentable.
Registration returns a typed manifest (per-leaf
path/dtype/shape/bytes, topology hash, determinism tier) that travels with
the moving object as its byte-exactness check. A native Rust data plane
(C-ABI, in-process) realizes the same critical path off the interpreter
(Table~\ref{tab:migcost}, \S\ref{sec:eval-m4}).

Admission itself is arithmetic. A deadline-class candidate is admitted iff
its measured per-SKU quantum, the resident state, and the guard band all fit the declared contract, computed from the deployment's own profile rows in place of model-family constants (full arithmetic, TR~\S E).

\looseness=-1{}After admission, a session holds its contract by degrading when slack runs
short. Each engine keeps a tier ladder, and the scheduler degrades to the
highest tier whose quantum fits the remaining slack. The degradation is
contractual and never silent.

\paragraph{Multi-GPU mechanisms.}\looseness=-1{} Five serving-layer mechanisms carry the
fleet results. Topology-keyed placement, prewarm shadow lanes, and
fail-closed migration commit are visible in Figure~\ref{fig:arch}
(profile rows are per-placement); block-boundary yield points (sessions
never preempt mid-quantum) and anti-thrash hysteresis (a move must save
more than it costs) operate underneath.

\paragraph{Implementation.}\label{sec:impl}\looseness=-1{}
\sys{}'s control and serving orchestration is Python: a per-node controller, one CUDA-pinned engine-host
process per GPU, Unix-domain sockets. Its migration data plane realizes the canonical
flatten/verify/commit path of \S\ref{sec:law-sufficiency} in-process, in two implementations reported
side by side in Table~\ref{tab:migcost}: a Python reference path and the native Rust C-ABI mover
introduced above. The full movable set is the
suspend manifest's $450$ leaves, the $420$ movable
tensors of \S\ref{sec:measure} plus scalar control flags. A fleet-scale
Rust control daemon is future work. The serving layer adds 0.2\,ms per quantum over the bare forward ($21{,}878$ quanta
dual-timed), nearly three orders of magnitude below the unit of work, with no
measurable cross-engine contention ($162.8$ vs $163.0$\,ms p50).
Telemetry is fail-closed. Every run fingerprints the host, asserts collector
liveness within $15$\,s, and aborts instead of producing unattributable
numbers. Both backends run unmodified except a $19$-line batch diff on
MG2's action module (it assumed batch size~$1$, walling off co-batching).
Under the same contracts, the artifact ships a playable single-GPU demo
at $27.8$\,fps on a consumer RTX~5090.

\section{Evaluation}
\label{sec:eval}
The evaluation argues one chain. Migration is deterministic within an architecture and breaks across
one (\S\ref{sec:eval-m5}). At memory speed and against active
generation, the primitive moves live state bit-exactly
(\S\ref{sec:eval-m4}). At session and population scale, the admissibility condition
instantiates as a measured envelope
(\S\ref{sec:eval-fleet}). At line rate the binding constraint moves
from transport to verification placement (\S\ref{sec:eval-xnode},
\S\ref{sec:eval-verif}). Finally, the
contract is grantable and placement-sensitive on the serving substrate
(\S\ref{sec:eval-substrate}).
The scope is bounded. Same-node and live-chase moves carry real MG2 state. With no live engine attached, the four-node line-rate cell moves the real state's bytes, and the two-node mover benchmark stays transport-level on synthetic payload. Over loopback,
the serving loop closes end-to-end at commodity rate, and fleet
placement beyond the executed 48-move loop is simulation. We have not
run the live serving loop over the $100$\,Gb fabric.

We evaluate on three node substrates: RTX 5090 workstations (the local measurement-study box of \S\ref{sec:measure} and a rented single-GPU instance for churn iteration), a rented A800-80G PCIe node, and a rented 8$\times$A100-80G SXM4 NVLink node. Together with an A100-SXM4-40G calibration probe these are the four GPU SKUs
of the fleet evaluation. With the H100 native-primitive node and the
L40S cross-architecture probe the paper spans six GPU SKUs across three
rental clouds (Vast.ai, AutoDL, Lambda) and an academic testbed (the
CloudLab d6515 line-rate fabric). Where a comparison is
claimed, the baselines are the two arms of our industry-default
module run under the same engine: (a) per-session GPU pinning
with FIFO queueing (pin-queue) and (b) deadline-blind round-robin
interleaving. The class-blind policies of \S\ref{sec:measure} are the
single-device ablations. (Deadline-blind greedy co-batching is ruled
out analytically by the measured coarsening curve of
Appendix~\ref{app:serving} and not run as a live arm.) Multi-GPU iteration results (\S\ref{sec:eval-m4}) are single-trial and
marked as such. The placement/determinism matrix (\S\ref{sec:eval-m5}) is
3-trial per cell, with tabled quantiles under one
definition (Hyndman-Fan q7). Every headline ratio is
recomputed by script from raw per-quantum ledgers archived with the
artifact. The workload suite is \emph{trace-calibrated}. Its demand is
parameterized from recorded traces and public statistics and \emph{replayed through the scheduler and the fleet economy}. Arrivals fitted from Azure LLM Inference~\cite{azurellm} and BurstGPT~\cite{burstgpt} drive admission, and per-class duty follows the same traces. Under that replay, the device-reclaim result and the reclaim economy both hold. No production world-model trace is public (\S\ref{sec:limits}). Demand follows a four-rung ladder:
controlled stressors (causal isolation); recorded open-oasis human play
(native human control, entropy $2.8$--$7.6$ bits); a \emph{gaming-session} rung fitting the
interactive class from $776$ desktop-gaming sessions (median $24$\,min;
input duty median $0.86$ with an AFK bottom decile, consumed directly by
\S\ref{sec:eval-fleet}); and a \emph{physical-AI} rung shaped from
published episode statistics.

\subsection{Determinism and the architecture boundary}
\label{sec:eval-m5}

This subsection locates the determinism tiers: bit-exact replay holds within an architecture at no enforcement cost and breaks across one (C4).

Same-seed sessions on different physical GPUs of the same architecture
produce bit-identical per-block latent streams ($12/12$ blocks, two
independent pairs; the surviving pair stayed bit-identical under
co-resident bulk interference). Fixed kernels on a fixed ISA make
reduction order reproducible, so T0 is a contract a scheduler can grant
at native rate, but only \emph{within} an architecture, which is why
migration beats re-derivation across one. Across the boundary
native-rate T0 is formally invalid (software-only emulation reproduces
the bits offline~\cite{bitexactverif}; no native-rate path does). We
measured where the decoded \emph{readout} then lands (A100$\times$H100
and A100$\times$L40S; Appendix~\ref{app:crossarch}). In every run the per-block cache CRC diverges at block~0, yet the
post-transient readout holds at the bounded tier T1.
\looseness=-1{}T0 is a property of a \emph{controlled configuration}.
Even within one architecture it breaks silently (a faulty P2P fabric,
measured in \S\ref{sec:eval-m4}; a transitively-downgraded cuDNN; a
co-batched pool that reorders reductions), while $9/9$
within-architecture controls hold on a pinned stack. Each break is
invisible to the output and caught only by the byte-identity check, a
second reason the primitive verifies bytes at the endpoints instead of
trusting the transport.

\subsection{The primitive on live state}
\label{sec:eval-m4}
\paragraph{Exhibit 1: end-to-end migration, certified bitwise, zero
added misses.}
\looseness=-1{}We ran an end-to-end migration cell on 2$\times$A800 in which a live
session advances on GPU~0, migrates to GPU~1 at a clean block boundary,
and continues there. With the destination engine pre-warmed (the
realistic fleet target), the migration itself is a $101$\,ms cache
install absorbed within one block's slack, with \emph{zero} added
deadline misses at a $600$\,ms contract. Two independent checks certify
bitwise continuity across the move, and a perturbed-cache negative control
diverges on every subsequent block (certification detail in
Appendix~\ref{app:churn}). One rented
host's direct-P2P path silently corrupted bytes mid-transfer and the CRC
check failed closed. No corrupt world was ever served. No link-level
check (fabric ICRC, PCIe LCRC) can see corruption on the host DMA
path, so byte identity is decidable only at the endpoints, the end-to-end
argument~\cite{saltzer84} measured at state-motion scale. A consistency-tolerant transport (one that
checks only that the bytes arrived, as production movers
do~\cite{turboserve}) installs such a corruption unflagged and serves a
wrong continuation, shown in a runnable head-to-head (artifact
\texttt{examples/demo\_migrate.py}).

\looseness=-1{}The fabric sets the cost of that certified move, and the
code path barely changes it. The Python path takes $18.8$\,ms end-to-end on
2$\times$H100 (continuation bit-exact against a no-migration reference),
a $16.6$\,ms Rust-FFI critical path against a corrupted-snapshot control,
and $83.9$\,ms on PCIe. Across those fabrics the move varies
$\sim$$5\times$ while the code path varies only $\sim$$1.1\times$
(Table~\ref{tab:migcost}), so the transport is commodity hardware and the
endpoint byte-identity guarantee is the contribution.

\label{sec:eval-crossmodel}
The same mover is model-blind. It moves a live Qwen2.5-0.5B~\cite{qwen} KV
cache ($24$ layers, $48$ leaves) in $4.03$\,ms round-trip, and greedy
decoding from the rehydrated cache is token-for-token identical to a
never-migrated reference ($16/16$ tokens, two independent runs; artifact
\texttt{mig\_llm}). The heavily-mutating regime that
\emph{needs} it is the world-model case (\S\ref{sec:related}).

Table~\ref{tab:migcost} traces the same move across the fleet's fabrics, with the cost keyed by the deployment's own profile, whatever the link advertises.
Block-boundary migration is effectively free on NVSwitch ($20.6$\,ms
modeled from the measured d2d rate, under one serving quantum), while a
virtualized host advertising \texttt{peer\_access=true} reaches useful
speed only when staged through pinned host memory.
\begin{table*}[t]\centering\footnotesize
\setlength{\tabcolsep}{6pt}
\caption{Migration-cost ledger, grouped by what varies: the code path (fabric fixed), the fabric (code path fixed), installation into a live serving cell, and the cold rebuild a move avoids.}
  \label{tab:migcost}
\begin{tabular}{@{}r l l l p{6.4cm}@{}}
\toprule
ms & software path & fabric & payload & includes \\
\midrule
\multicolumn{5}{@{}l}{\emph{Code path varies, fabric fixed}}\\
16.6 & Rust-FFI flat-pack & 2$\times$H100 SXM NV18 & 1.67\,GB/420t & critical path (pack/wire/host/unpack stages overlapped); corrupt-control \\
18.8 & Python path & 2$\times$H100 & 1.67\,GB/450 leaves & flatten+\texttt{cudaMemcpy}+CRC+rehydrate ($101\times$ vs $1899$ save/load) \\
\addlinespace
\multicolumn{5}{@{}l}{\emph{Fabric varies}}\\
83.9 & Rust-FFI flat-pack & 2$\times$A800 PCIe & 1.67\,GB/420t & same cell as the 16.6 row \\
71.1 / 96.4 & raw P2P & 8$\times$A800 intra/cross-island & 1.67\,GB & wire only \\
\addlinespace
\multicolumn{5}{@{}l}{\emph{Into a live serving cell}}\\
101.2 & warm install, live cell & 2$\times$A800 & 1.67\,GB & serving-cell install, 0 added miss \\
74--124 & fleet-executor warm & 4$\times$H100 SXM5 & 1.67\,GB session & snapshot+wire+install (\S\ref{sec:eval-fleet}) \\
\addlinespace
\multicolumn{5}{@{}l}{\emph{The rebuild a move avoids}}\\
$\sim$$11{,}200$ & first-touch cold & 4$\times$H100 SXM5 & --- & lazy engine materialization (prewarm target) \\
\bottomrule
\end{tabular}
\end{table*}
Cold reopen (no state
transfer) measured 260--322\,ms across hosts. An 8-engine churn run on
the NVSwitch host reproduced the tuned scheduler behavior (admission estimate within $1.1\%$).
\paragraph{Exhibit 2: migration against a generating session.}\looseness=-1{} The
primitive's hardest case is a live session, where the
model rewrites its cache while the mover chases it. We ran exactly that
on two H100s with the shipped iterative pre-copy API. Pre-copy cannot
converge while the model generates (the per-round dirty set saturates at
$0.777$ of the cache) and converges the moment generation pauses at a
block boundary, with $1.14$\,s fail-closed downtime (stepwise chase in
Appendix~\ref{app:churn}). The destination then \emph{continues the
world}. Its next three generated blocks are bitwise identical to the
source's counterfactual continuation. To our knowledge this is the first
bit-exact live migration of an actively generating world-model session.
Convergence is $\rho_{\mathrm{bw}}{=}D/B$ with $D$ scaling by step rate.
On this wire at H100 rate $D{>}B$ and pre-copy saturates. On a
$100$\,Gb fabric the dirty set follows a \emph{zero-parameter} coupon
law whose calibrated recursion \emph{predicted} the convergence boundary
before the run and held (Appendix~\ref{app:churn}), so the
safe-churn region is measurably wider than the classical $D{=}B$ line. The
divergence threshold $D^{*}$ inherits its $B$ from the verified-rate
composition (\S\ref{sec:eval-verif}), so the boundary is placement- and
engine-dependent. The dirty-flux term of the condition is measured on both
sides here, saturation while generating and convergence at the block boundary.

\looseness=-1{}Across 15-minute churn runs the scheduler tuning ladder took C1 misses
from $100\%$ to $0.69\%$ while bulk went from starved to $2{,}519$ blocks
(admission's service estimate within $2.5\%$ of measured). Planner bursts
completed $52/52$ branches at steady makespan.

\subsection{The operating envelope and the population cell}
\label{sec:eval-fleet}

\looseness=-1
\textbf{Exhibit 3: the condition becomes an operating envelope.} On the
multi-GPU nodes the admissibility condition instantiates as a two-axis
region. A deadline gate (frozen $33$\,Gbps, live $41$\,Gbps, both at
$400$\,ms) sets the feasibility floor; above it, prewarm lane count sets
the feasible oversubscription. Fabrics at $10$ and $25$\,Gbps stay
inadmissible at any lane count for raw motion, while above the gate $\nu$
rises with prewarm lanes toward the Engset cap ($2.95$ at ${\ge}100$\,Gbps; all
per-fabric values are projections, TR~\S E). At fixed
state the gate is aging out. At the $800$\,Gb/s per-GPU NICs now
shipping~\cite{cx8}, this cache's wire time is $\sim$$17$\,ms, the
binding constraint moves from link bandwidth to \emph{per-destination}
serialization (the incast row of the fleet test), and placement becomes
the scheduled resource. Against growing state the gate re-tunes instead
of retiring. The minimum admissible bandwidth $C/H$ is linear in state,
and at the measured dirty fraction the $1.67$\,GB cache needs
${\approx}41$\,Gbps where a $32$\,GB cache needs ${\approx}783$\,Gbps,
almost exactly one of those $800$\,Gb ports (size-ladder fit in the
artifact). The consolidation loop below executes inside the region,
and the cross-node mover crosses its gate at line rate
(\S\ref{sec:eval-xnode}). At a production $10$\,Hz step rate
($D{\approx}13.5$\,GB/s) the dirty-dependent boundaries rise to
$\approx$$141$ and $\sim$$108$\,Gbps.\label{sec:law-corollary}
\looseness=-1{}A four-node CloudLab cell (4$\times$d6515, one $100$\,Gb RoCE
LAN, \$0) measures that incast row directly. At $\sim$$73$\,Gb/s, the solo $1{\to}1$ migration of the real
$1{,}670{,}124{,}960$-byte state runs in $182$\,ms,
bit-exact. The cell's mover runs below the $92$--$95$\,Gb/s
tail-scheduled anchor of \S\ref{sec:eval-xnode}, both clear the gate, and
the collapse below is measured against the cell's own solo rate. Piling
$K$ movers onto one $100$\,Gb receiver collapses per-move
goodput and drops the receiver's aggregate ingest below the solo rate
(Table~\ref{tab:incast}). The onset is at fan-in $K{=}2$, and total ingest
falls where fair sharing would have held it near $73$\,Gb/s, a super-linear
wall of $1.8$--$3.2\times$ the clean-sharing bound. The collapse is silent. Every completed move is bit-exact ($49/49$ sent moves, CRC and byte count
verified, zero transport failures), so what breaks is the SLA, the discrete
second-scale freeze under incast the admission model already predicts, here
measured at population scale. Spread across receivers, the same movers stay clear of the collapse: two
movers to two receivers hold $73.0$\,Gb/s ($0/6$ miss) and disjoint pairs
stay flat at $182$\,ms, while a single sender's egress fan-out runs at line
rate ($101$/$103$\,Gb/s aggregate at $K{=}2$/$3$). The pathology is
receiver-ingress incast specifically,
so the cell reset the controller's receiver model. Multi-source fan-in is a
different regime from same-host dual-flow, so the controller now treats a
multi-source receiver with one move in flight as at capacity. This
transport-incast onset ($K_{\mathrm{inv}}{=}2$) is a separate mechanism
from the verification-plane crossing $K^{*}{\approx}4.2$ of
\S\ref{sec:eval-verif}. At this cell's maximum fan-in of $3$ the
fair-shared wire stays below the CRC bound, so the crossing is out of
reach by arithmetic and the cell validates the transport collapse and
the receiver-spread fix. Reaching $K^{*}$ needs ${\ge}5$ nodes.

\begin{table}[t]\centering\footnotesize
\caption{Single-ingress incast on the $4$-node cell: $K$ full-cache movers
  piled onto one $100$\,Gb receiver. Per-move goodput collapses and the
  receiver's aggregate ingest falls below the solo rate; every move stays
  bit-exact.}
\label{tab:incast}
\setlength{\tabcolsep}{5pt}
\begin{tabular}{cccc}
\hline
fan-in $K$ & per-move & into receiver & on-time \\
           & (Gb/s)   & (Gb/s)        &         \\
\hline
1 & 73.0 & 73 & yes \\
2 & 20.4 & 43 & no  \\
3 & 7.8  & 25 & no  \\
\hline
\end{tabular}
\end{table}

\paragraph{An incast-aware controller restores the zero-miss invariant.}
A naive pair-calibrated controller (min-queue-depth picker on the optimistic
$\mathrm{max\_fan\_in}{=}2$) admits into the collapse. Of $72$ offered moves
it lands $10$ on time, misses $57$, and rejects $5$, with admitted goodput at
a $4.4$\,Gb/s median ($0.79$ miss fraction). At the same load, the incast-aware controller
(fan-in cap $1$ and receiver-spread, with a capacity-aware pick over the
measured per-path rate matrix) lands $35$, misses $5$, and rejects $32$,
and holds admitted goodput at a $72.7$\,Gb/s median, matching
the solo rate. This is a $0.07$ miss fraction, $11\times$ fewer misses and $3.5\times$ more
on-time moves. Admitted misses stay at zero up to $1.4\times$ offered load,
and past that overload exits as an explicit reject, the load-shed the
admission invariant specifies. At $2\times$ load a $5/72$ residual miss
remains, each one a genuine receiver-ingress collapse ($11$--$38$\,Gb/s), so
placement narrows the shared-ingress capacity wall without hiding it under
sustained overload. Because the harness generates each payload before sending, it paces
admission decisions cleanly but the wires only approximately. That gap is
the likely reason the cap holds to $1.4\times$ and gives way at
$2\times$; tying wire-start to decision time (a resident-state cell)
would close it. Rate heterogeneity is sender-side
(one slow egress node at $54$\,Gb/s) with uniform receivers, so the naive
$57/72$ miss was pure incast.

\looseness=-1{}This collapse and the scheduled contention law are the two
ends of one receiver-serialization axis. $T(N)=250+165(N-1)$\,ms
(measured $10/10$ bit-exact under contention, within $6\%$ of the model
with verification charged) is the same $100$\,Gb receiver serving $N$ movers
one at a time ($T(2){=}415$\,ms), and the barrier-synchronized incast here is
the unscheduled case where all $K$ land at once (per-move $\sim$$655$\,ms at
$K{=}2$, $\sim$$1.7$\,s at $K{=}3$). The incast-aware controller converts the
unscheduled collapse into that scheduled serialization, so the interactive
fix for receiver-ingress incast is receiver-spread. Lossless compression is a
separate handle, measured in \S\ref{sec:eval-xnode} as a
gate-widening option and a byte saving on the evacuation and bulk budgets.

\ifextended
Specialized to the $1.67$\,GB MG2 cache, the condition places the measured
systems on one map (Figure~\ref{fig:regimemap}).

\begin{figure}[t]
  \centering
  \includegraphics[width=0.79\columnwidth]{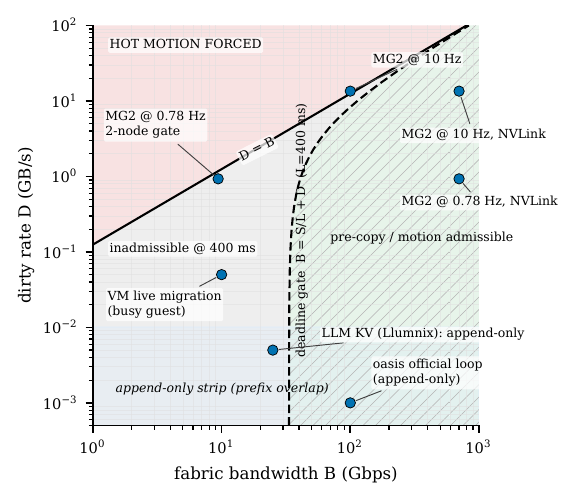}
  \caption{The serving-regime map: each object a point (dirty rate $D$
  vs bandwidth $B$). Left of the gate migration misses the deadline;
  above the $D{=}B$ phase line pre-copy cannot converge. MG2
  ($\rho_{\mathrm{WWS}}{\approx}0.75$) crosses regimes as step rate
  scales $0.78\to10$\,Hz; append-only objects never leave.}
  \label{fig:regimemap}
\end{figure}

\looseness=-1{}The map's left edge is the live deadline gate of
\S\ref{sec:law-fleet}; the flux clause becomes the \emph{dirty-rate phase
line}, $D{=}B$ at $\sim$$7.5$\,Gbps for the measured $D{\approx}0.93$\,GB/s.
Both dirty-dependent boundaries scale with step rate, so the map is the operator's admission chart. Locate an
object's $(D,B)$ point, and the region names its legal primitive.
\fi

\looseness=-1{}The envelope so far sizes one object's admission. At fleet scale the resource is a population of these objects, and the question turns to what consolidation can reclaim from it. Under the strict-divergence contract only two safe ways to free a GPU remain. \emph{Migrate} the live state (warm install $96.4$--$101$\,ms, constant in session age), or \emph{evacuate and replay} at $O(T)$ forward passes to rebuild step $T$. Any long-lived world is therefore dominated by constant-cost migration, the only history-independent primitive we can certify (pre-copy convergence: \S\ref{sec:measure-dirty},
\S\ref{sec:eval-m4}). To price the fleet economy it buys, and where it
buys nothing, we model $N$ sessions as
\textsc{on}/\textsc{off} processes, pin-per-user versus a migrate-on-idle
pool (evacuate past a keep-warm threshold, migrate back on reactivation),
reporting the minimum pool size holding a $1\%$ reactivation SLO-miss
bound. ``Idle'' means the session is \emph{suspended}: the forward pass has
stopped and the cache is frozen. A session idling at the keyboard is
still live, because the model still rewrites its cache every step
(\S\ref{sec:measure-dirty}; population examples,
Appendix~\ref{app:envelope}). The simulation is driven by the \emph{measured}
migration cost. Code and manifests ship in the artifact.

\paragraph{The consolidation loop executes on real hardware.}\looseness=-1{} We close
the loop in \emph{execute} mode. Every migration the rebalancer schedules is
performed as a real WorldMove (snapshot$\to$transfer$\to$verify$\to$install)
rather than charged from a cost model. On a 4$\times$H100 node, three churn seeds
reproduce the modeled scheduling trajectory exactly (the rebalancer is deterministic, so what execution adds is that every move is demonstrated feasible and timed on hardware), and the walls expose the hot/cold boundary in-band.
Warm installs take $74$--$124$\,ms and first-touch migrations $11.0$--$11.5$\,s of
lazy initialization, observed inside a live loop.
At $N{=}8$ on two independent providers' NVLink nodes the executed and
modeled trajectories again match seed-for-seed ($2.263$ devices freed over
$11$ migrations, identical on both providers), and idle-reclaim (null at $4$ devices) appears at $8$, so the reclaim population depends on fleet size. One schedule
was refused fail-closed by VRAM admission the capacity-blind simulator had approved. In total,
\textbf{$48/48$ executed migrations verified bit-exact} across two fleet
scales, three nodes, and two scheduling modes.

The yield is a pooling economy that scales with fleet size and demand
mix, and its shape follows from the measurements. First, it is \emph{prewarm-conditional},
with break-even prewarm hit-rate $\lambda^{*}{\approx}0.64$\ifextended{} ($25$--$75\%$
band $[0.55,0.73]$, $64$ seeds; hot copy $52$--$152$\,ms vs
$11{,}974$\,ms first-touch, $79$--$229\times$)\fi. Because measured idle
windows are $4$--$24\times$ shorter than the $\sim$$12$\,s re-init, the
economy runs on \emph{residency}, the world already warm on the
destination before the pause begins. Second, it is a
\emph{region}: hot pools pass on every seed, cold pools on none. Finally, the
interactive \emph{duty} driving it is derived from traces.

The sizing itself is classical finite-source loss (Engset;
\S\ref{sec:related}). Only the primitive being sized is new.

\looseness=-1{}Bit-exact deterministic state
motion is the enabling condition for tail-SLO consolidation. With
think time $Z$ between a session's generation bursts, sessions-per-device
is that finite-source machine-repairman problem~\cite{scherr}
(setup-cost farms,~\cite{gandhi}), instantiated with measured cost constants
(Fig.~\ref{fig:kmax}, Appendix~\ref{app:envelope}). The host-swap region exists only because
restore is \emph{deterministic}, and that is the non-classical part. With exponentially distributed restore at the same
mean, the $400$\,ms/$1\%$ SLO is infeasible at \emph{any} $K$ (the
own-service tail alone is $7.4\%$). On the
measured cost constants an $11$\,GB/s host tier lifts an $80$\,GB device from
$40$ resident sessions to $K^{*}{=}46/181/902$ at $Z{=}30/120/600$\,s
($1.15$--$22.6\times$ residency, Fig.~\ref{fig:kmax},
Appendix~\ref{app:envelope}\ifextended; a consumer PCIe-5 host comes in
$4.4\times$ under the model, and in the interference arms contention
perturbs timing but never bytes, Appendix~\ref{app:churn}\fi). The
long-$Z$ cells cover the persistent-engine regime this consolidation
targets. MG2 itself drifts to unusable within ${\sim}10$\,s of
continuous generation (\S\ref{sec:limits}), so the $Z{=}600$\,s cell
is a fleet-class projection.

\paragraph{Scope.} Fleet numbers are simulation calibrated by measured
constants; the loop itself executes on real hardware above.

\looseness=-1{}Trace-calibrated validation (Azure LLM-inference~\cite{azurellm} and
BurstGPT~\cite{burstgpt} arrival processes replayed against the measured
constants) reproduces the same hard-zero and reclaim-region structure
end-to-end.

The reproduced hard zero at duty${\to}1$ is itself a property of the assumed interactive duty, and measured duty moves it. Feeding the gaming rung's per-session duty
distribution (\S\ref{sec:eval}) into the finite-source sizing reclaims
$13.2\%$ of the interactive fleet at a $5$\,s suspend threshold against
$0.8\%$ under the synthetic duty-$0.95$ assumption\ifextended{} ($6.1$--$19.3\%$
across thresholds; a majority-interactive fleet reclaims $22.5\%$ gross,
Fig.~\ref{fig:migphase}, Appendix~\ref{app:envelope})\fi. These are gross
ceilings, and the warm-buffer break-even is in Appendix~\ref{app:envelope}.

\subsection{Line rate}
\label{sec:eval-xnode}
\textbf{Exhibit 4: the gate is crossed, and verification placement
binds.} \looseness=-1{}On the $100$\,Gb fabric the Rust mover completes the
CRC-verified $1.67$\,GB move cross-node in $141$--$146$\,ms median
($92.1$--$94.8$\,Gb/s against the mover's $97.95$\,Gb/s verify-off
ceiling), inside one $400$\,ms interactive block and above the
$41$\,Gbps live gate. What remains to schedule at that rate is where the
verification runs. Below, the first-generation mover's $251$\,ms median
(bit-exact, nine trials, eight warm) is retained as the
software-floor exhibit. Two real two-instance moves, a $25$\,Gb RoCE
tier, and the $100$\,Gb fabric below are the measured anchors. A
\texttt{netem}-controlled sweep maps how latency scales with bandwidth
between them, and rates we could not rent remain labeled projections.

Two of those anchors are moves over rented inter-node fabrics, both
bit-exact: two
physical H100 instances at their $\sim$$9.4$\,Gbps NIC ceiling and an
earlier A10 pair at $\sim$$2$\,Gbps, giving two hardware $B$-points for
the fleet test of \S\ref{sec:law-fleet}, both below the $41$\,Gbps gate
(path detail: Appendix~\ref{app:envelope}).
\looseness=-1{}The $100$\,Gb anchor is two CloudLab
d6515s. With verification off, the Rust mover measures $97.95$\,Gb/s,
the verbs ceiling itself ($\sim$$0.1\%$ mover overhead); verified, it
holds $92.09$\,Gb/s at uniform $64$\,MB chunks and $94.8$\,Gb/s with the
tail schedule (Fig.~\ref{fig:placement}, Appendix~\ref{app:swfloor}). At both chunk sizes measured, the
exposed verification tail equals exactly one final-chunk CRC
($21.7$\,ms at $256$\,MB, $8.2$\,ms at $64$\,MB), so the verified rate
is wire $\times$ wall/(wall $+$ one chunk CRC) and the tail is a
scheduling choice. Sequential placement measures $80.22$\,Gb/s.
The Python transport that preceded it is the software-floor exhibit.
Its \emph{CRC-verified end-to-end $251$\,ms median} move
($n{=}8$ warm trials, max $301$\,ms) carries $53$\,Gbps verified
(registration and CRC-engine forensics:
Appendix~\ref{app:swfloor}). Over loopback, the serving loop itself is
closed end-to-end: a live session generated $12$ blocks,
suspended at a block boundary, shipped its $1.67$\,GB state through this
verified mover (manifest-bound pieces, receipts on both sides), restored
fail-closed, and continued \emph{bit-exactly} against a no-migration
control. Across $3/3$ seeds the match held on five equality gates,
including the $420$-leaf tree and the four-block continuation trajectory,
with the first block after restore within $10\%$ of the control median. Below line rate too, the floor is
removable layer by layer (the $25$\,Gb RoCE tier and the \texttt{netem}
sweep: Appendix~\ref{app:swfloor}).

A floor sits under that measured rate.
Under the exact-continuity contract a legal primitive must materialize
the source state's exact bytes at the destination. Against a destination
that holds no prior replica, \emph{raw} byte motion on a path with
achievable rate $B$ is bounded below by $S/B$. On this fabric that floor
is the $97.95$\,Gb/s verbs ceiling above, $\sim$$136$\,ms for the
$1.67$\,GB cache. The raw floor is not the information floor: a
contract-legal codec ships below it, measured next. Each evaluated alternative
either fails the contract or sits above the floor: every lossy handle in
the kill-family breaks $R$ (\S\ref{sec:measure}), cross-architecture
recompute diverges at native rate at the first block
(\S\ref{sec:eval-m5}), replay from history grows with session age and
outruns the horizon (the motion criterion, \S\ref{sec:law}), and
save/load, the only other deployed exact-class primitive we know of, measures $101\times$
slower (\S\ref{sec:eval-m4}). The verified rate above therefore sits
within $6\%$ of this raw mover's own verify-off ceiling. That codec is
lossless compression, the remaining reduction handle against a cold
destination. Whether it pays turns on where the codec runs.
On the real warmed cache ($99.99\%$ bf16, byte entropy $5.87/8$\,bits) a
bit-exact quotient removes up to $\sim$$40\%$ of the bytes whole-tree
(CPU zstd-19), but the fastest CPU codec, lz4 at $19\%$ removed, costs
$489$\,ms to compress and decompress the $1.67$\,GB payload, past a full
serving quantum ($148.5$\,ms). On the GPU the same quotient is admissible:
nvCOMP's ANS entropy coder removes $35.6\%$ whole-tree ($44.6\%$ with a
bf16 byte-plane split) at $230/252$\,GB/s, a $\sim$$7$\,ms compress step
$20\times$ under the quantum, every round trip bit-exact. The reduction
concentrates in the redundant parts: the readout-irreducible world-state
KV compresses only to $0.80$, near entropy-saturated, while the action
cache ($0.40$) and the bf16 exponent plane supply most of the blend. The
receipts bind to the decoded tree, so the codec sits inside the verified
envelope. We cost the option from the measured codec and wire and leave
the integrated compressed move unbuilt. (In place, the same codec shrinks
a suspended cache by the same fraction as a capacity discount, but it frees
no device.) The codec's value is a wider admissible region. At line
rate it saves only $\sim$$34$\,ms, but netting codec against wire widens
the deadline gate of \S\ref{sec:law} by the compression ratio, carrying the
$25$\,Gbps fabric from infeasible for raw motion ($553$\,ms) to feasible
($\sim$$370$\,ms). Below the gate, where the
single-ingress incast of \S\ref{sec:eval-fleet} collapses the per-move
rate to $20$ and $8$\,Gbps, the same reduction saves $\sim$$220$ and
$\sim$$600$\,ms, but on moves already past the interactive horizon, so
that wall time is an evacuation and bulk-relocation budget that the live
contract never claims. At this layer what stays contestable is legality and verification
placement. For this object class, transport speed is no longer the contested resource. Commodity KV movers ride the same
wire and none of those we examined check the contract end to end. The codec's decode raises the
same receive-path question as the CRC, and where that check runs is a
placement problem.

\subsection{The verification plane}
\label{sec:eval-verif}
\begin{figure}[t]
  \centering
  \includegraphics[width=\columnwidth]{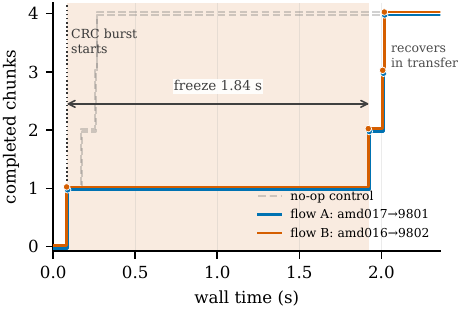}
  \caption{Two concurrent full-cache verified moves through one receiver
  (barrier-synchronized incast, $512$\,MB chunks): in collapsed
  repetitions both flows freeze \emph{synchronously} for
  $1.7$--$1.8$\,s; the no-op-executor control is stable.}
  \label{fig:freeze}
\end{figure}

The verification plane has three regimes. First, at these rates verification is a second scheduled resource, no longer a fixed overhead. Coupling host
checksumming to receive processing goes back to the earliest TCP
overhead analysis~\cite{clarktcp89}, ZFS-class end-to-end checksums are the
storage-plane precedent, and the GridFTP lineage overlaps checksum with
transfer to hide exactly this cost~\cite{fiver,liupipeline}. What is new
at $100$\,Gb kernel-bypass state motion is the failure mode. The
placement that hides the cost trips a
protocol-timer-quantized stability failure, so
verification enters admission control itself. Overlap alone cannot place it safely. The verified rate composes as
$\mathrm{harm}(B_{\mathrm{wire}},B_{\mathrm{crc}})\le
B_{\mathrm{verified}}\le\min(B_{\mathrm{wire}},B_{\mathrm{crc}})$, and both
bounds are met in measurement (sequential: $53.2$ vs harmonic $53.4$\,Gbps, $0.3\%$; overlapped, the $21.5$\,Gbps RDMA path of Appendix~\ref{app:swfloor}: $93\%$ of $\min$). A bit-exact mover's ceiling is therefore whichever engine saturates first. Hiding the checksum inside the wire
(per-chunk CRC overlapped with receive, GF(2)-combined) removes the CRC term from the critical path. Solo, verified e2e drops from $251$ to $215$\,ms (sender frame) and the measured cost reduces to $T=\mathrm{wire}+c$, $c\approx55$\,ms across warm trials ($54$--$65$\,ms under fan-in), the same wire-plus-final-verify shape the one-chunk-tail law gives on the Rust mover.
But the two planes couple, and $2{\to}1$ fan-in amplifies the coupling
into a stability failure. In $2/3$ of the barrier-synchronized
incasts~\cite{incast} the wire enters discrete second-scale freezes, and
matched controls isolate the CRC work as the trigger
(Fig.~\ref{fig:freeze}; Appendix~\ref{app:forensics} holds the episode
anatomy, the controls, and the counter-level inference chain).

 The
freeze's constant component \emph{is} the retransmit timeout,
hardware-quantized at $4.096\,\mu\mathrm{s}\times2^{n}$ (the verbs
local-ACK-timeout quantum~\cite{ibta}). Quartering the configured
timeout shifts the constant $2147{\to}536$\,ms ($4.006\times$) yet
prevents no episode, the NIC's minimum timeout exponent clamps further
tuning, and across $4.3$ hours of concurrent probing the regime stayed
invisible from above ($14{,}973$ ICMP samples, max RTT $0.441$\,ms,
zero loss) while the bulk plane sat in multi-second freezes. As
configured, a drop is repaired by go-back-N retransmission, not
prevented by PFC or paced away by DCQCN. The loss literature
(DCQCN~\cite{dcqcn}, RDMA at cloud scale~\cite{rdmascale},
IRN~\cite{irn}) treats go-back-N loss as congestion to engineer away,
while the loss here is induced by receiver-side verification compute, a
trigger that literature does not cover.

Second, checksum capacity caps the fan-in.
A receive path that checksums at $B_{\mathrm{crc}}$ sustains fan-in of
$K$ wires of rate $B_{\mathrm{wire}}$ only while
$K\,B_{\mathrm{wire}}\le B_{\mathrm{crc}}$, so receive-path verification
stops scaling at $K{=}B_{\mathrm{crc}}/B_{\mathrm{wire}}$: the measured
$32$-thread engine ($52.6$\,GB/s of CPU-side zlib CRC, $\sim$$2.4$\,GB/s
per core) against $100$\,Gb wires ($12.5$\,GB/s) crosses at
$K\approx4.2$, equivalently a $\sim$$420$\,Gbps single-wire crossover
($\sim$$120$\,Gbps for the per-flow CRC). The measured freeze occurs well below that bound
(arithmetic: Appendix~\ref{app:forensics}): the failure is a
transient, a CRC burst overlapping an active receive and tripping a
retransmit timer whose floor on this NIC ($536$\,ms) already exceeds the
$400$\,ms block. At $100$\,Gb the constraint that binds verification
placement is therefore timer-scale interference, with the capacity
crossing as the outer bound no receive-path engine at these CPU-side CRC
rates can beat. At $400$\,Gb/s the outer bound applies directly:
$B_{\mathrm{wire}}{=}50$\,GB/s puts the crossing at $K\approx1$ for this
engine, so a $52.6$\,GB/s CPU-side check cannot sustain even two-wire
fan-in and verification must leave the receive path (a NIC/DPU offload or
post-staged verification) on capacity grounds alone. The escape is a
faster check. Hardware-accelerated CRC (a PCLMULQDQ/AVX-512 fold of the
same polynomial) runs several times faster per core, raising
$B_{\mathrm{crc}}$ and moving the crossing outward, so the prediction is
scoped to the receive-host rate: a $400$\,Gb deployment whose CRC stays
below $\sim$$50$\,GB/s will fail under sustained fan-in even with the
timer pathology engineered away.

\looseness=-1{} Contention in a kernel-bypass
fabric is therefore still arbitrated, by the lowest
protocol timer that catches it and at that timer's clock; scheduling
chooses \emph{which} clock, $250$\,ms colored rounds instead of a doubled
retransmit timeout ($\sim$$2$\,s). The break-even is arithmetic:
overlap saves at most the hidden checksum time ($36$\,ms here;
$\propto1/B$ only once the wire outruns the CRC engine) and risks
$p\cdot E$ at protocol scale. It improves the \emph{median} and worsens
the \emph{mean} ($421$\,ms overlapped vs $264$ sequential), and faster fabrics make
in-path compute strictly worse. Verification \emph{placement} is therefore a stability decision rather than a
latency knob. On this stack, transport and verification must be admitted
\emph{jointly}. A NIC/DPU checksum offload would remove the host compute from the
receive path, the natural escape hypothesis, untested here. A
lossy-tolerant KV mover, which may simply skip verification, never faces
any of this. Composed with the fleet test, these numbers settle admission. At $100$\,Gb and $H{=}400$\,ms, sequential verification
misses $N{=}2$ ($405$--$409$\,ms); overlap admits it in the fast mode
($322$\,ms) and misses it in the stall tail (trip rate $2/3$--$12/12$ across hours), while the measured $32$-thread
engine re-admits it cleanly. At datacenter rate the admission integer
$N_{\max}$ depends on verification
placement and the checksum engine; bandwidth alone does not set it (raw records in the artifact).
The stakes of placement are a measured band of state sizes. At
$H{=}400$\,ms, $3.60$--$4.46$\,GiB is servable only with off-path
pipelined verification (frontier figure: TR~\S E). The $1.67$\,GB cache
sits inside the feasible region, so for this object placement is an
optimization; for caches in that band it is feasibility itself. No deployed mover we surveyed ships that capability: NIXL~\cite{nixl}
and Mooncake~\cite{mooncake} carry no end-to-end integrity check
(source-verified 2026-07), and where verified transfer exists it is a
file-transfer service that overlaps or post-stages its checksum off the
GPU fabric~\cite{fiver,liupipeline}. Off-path
pipelining itself stayed stable in every solo run on this mover ($0/48$
episodes), consistent with the freeze being a property of in-path designs.

The third regime is non-fluid. \looseness=-1{} Concurrent
full-cache moves through one switch measure two fluid placements and one exception. The fluid pair matches the
prediction: node-disjoint flows are free ($248$\,ms $=$ solo) and flows
sharing a destination port fair-share ($415$\,ms $\approx$ $2\times$ the wire share plus one CRC). The
exception: \emph{duplex} placements (one node sending and receiving
full-cache moves simultaneously) are winner-take-all, one direction at
line rate, the other at $1/5$--$1/10$ of it. Five of six mesh rounds show a
single rotating winner and one shows two. In $6/6$ two-node duplex runs
the winner is sticky by direction, independent of start order. Raw verbs
duplexes run clean at $195.5$\,Gb/s aggregate on the same pair, which
scopes the facet to the two-process mover stack. One admission rule follows:
\emph{budget flows per destination port, never co-schedule full-cache duplex pairs, and
place the checksum where fan-in cannot couple to it} (receive
livelock~\cite{mogul97} in kernel-bypass form: receiver compute starving
the receive path). The rule is
constructive. As two colored rounds of node-disjoint matchings (the
edge-coloring vocabulary of matching-based transfer scheduling), the same
four all-duplex transfers that free-run to a $2.0$--$2.4$\,s worst flow
complete in $539$--$545$\,ms. Predicted before the run, the coloring
bound (rounds $\times\,T_{\mathrm{solo}}$) measured within $9\%$
($3/3$, $16/16$ bit-exact). The $800$\,Gb corollary below inherits
these facets; link bandwidth alone does not set them.

\paragraph{The admission policies under load.} On the same pair, with
every admitted move physically executed and CRC-verified, two doctrines
without admission (an overlap-everything mover in the FIVER style and an
accept-all rate-limiter) collapse past capacity (miss fraction $0.94$ at
$1.4\times$), while the model-based controller matches a defer-and-reject
operator heuristic with no hand tuning: zero deadline misses at every
offered rate, overload absorbed as explicit rejections. Constants decide
it. Fed idiom-stale constants, the same controller missed
$0.56$ of nominal load, so constants must come from the deployment's own
execution idiom. A Jacobson-style estimator over the controller's own
completed moves reaches the same served fraction with no calibration
step. On a second fabric the comparison replicates, and a three-seed
replication on a third, re-imaged pair holds the invariant in all $24$
cells, zero deadline misses and zero errors, under synthetic arrivals to
a single destination (transplant caveats:
Appendix~\ref{app:envelope}).

\paragraph{Datacenter projection.} Subtracting the removable floor, the
line-rate wire cost is $134/67/33$\,ms at $100/200/400$\,Gbps; adding back
the $\sim$$19$\,ms native install floor as a fabric-independent constant, a $400$\,Gbps install lands near $52$\,ms. These are projections: the measured affine law plus the measured floor, extrapolated past the fabrics we could rent.

\paragraph{Snapshot \emph{and} live pre-copy.} Both modes are supported,
and the $\rho_{\mathrm{bw}}$ law stated whole in
\S\ref{sec:eval-m4} decides which converges.

\paragraph{A deployable decision.} \looseness=-1{}Where a deployment lands is a bandwidth
(gate) $\times$ prewarm (magnitude) operating point, and the five decision rules (deadline gate, pre-copy convergence, fleet schedulability, migration
economy, and verification placement under the measured capacity facets) compose to one decision per (fabric, SLO): migrate, prewarm, or reject. TR~\S B holds the
full \{fabric\}$\times$\{SLO\} grid, composed by \texttt{scripts/decide.py}
($B_{\min}{=}S/H{+}D$ from the measured constants): budget
fabric is rejected in every cell, NVLink admits at every SLO, and PCIe-RDMA
admits only at $\ge$$400$\,ms.

\ifextended The live serving loop and the verified line-rate mover have not
yet been composed on one $100$\,Gb fabric; that run is scheduled, and
\S\ref{sec:limits} states the boundary it closes.\fi

\subsection{Serving substrate}
\label{sec:eval-substrate}

The primitive runs on a serving substrate whose placement, policy, and quality-knob behavior we characterize here.

All numbers in this subsection are from 3-trial cells on an
8$\times$A100-80G SXM node unless noted. Demand is controlled-stressor
synthetic (for causal isolation) plus a recorded-player-trace replay arm
(TR~\S E).

\paragraph{Quanta are placement-sensitive.} An engine pinned to the
wrong NUMA node runs every forward at $\sim$$202$ instead of
$\sim$$152$\,ms, a $33\%$ penalty from PCIe/NUMA affinity alone,
and that is enough to turn an otherwise-feasible
$600$\,ms contract into a $\sim$$90\%$ miss rate. The failure is
deterministic, and re-running on the wrong node reproduces the miss. \ifextended Placement is fixed when the process spawns, and correct binding is necessary
but not sufficient (\S\ref{sec:limits}). A profile must therefore key to
the placement itself, because the same SKU spans both sides of the
deadline. The penalty splits by class: the same slow placement that lets
soft-deadline planner bursts finish $38/40$ makes hard-period
interactive anchors miss $77.8\%$.\fi

The policy's own effect is next. With every
arm NUMA-bound (8$\times$A800-80G split 4+4, 10-min cells, 3
trials, fresh process per cell; \sys{} and pin-queue honor the
2-step quality floor), the cross-arm comparison is causal. \sys{} holds C1 miss at $0.44\%$ while
completing the bulk job on every device. Pin-queue isolation matches the
miss rate ($0.36\%$, same quality floor, same frame-skips) but serves
\emph{zero} bulk. Blind round-robin completes the bulk but drives C1 to
$6.27\%$ ($14\times$). Isolation buys no QoS over \sys{}, which matches a dedicated GPU within per-device noise. Admission does most of the deadline job. The contract scheduler's marginal value concentrates in the warm transient, the degraded-latency frontier, and bulk goodput under guard.

\begin{table}[t]\centering\footnotesize
\caption{Placement-controlled policy comparison (1 interactive C1 + a
finite bulk job per device; per-cell figures). The blind baseline runs
C1 unprioritized at the full 3-step tier (residual confound: \S\ref{sec:limits}).}
\label{tab:abc}
\setlength{\tabcolsep}{3.5pt}
\begin{tabular}{lrrr}
\hline
policy & C1 miss\% & bulk blk/cell & C1 skips \\
\hline
\sys{} (EDF+guard+degr.) & 0.44 & 960 & 213 \\
pin-queue (isolate)          & 0.36 &   0 & 212 \\
blind-RR (naive)             & 6.27 & 960 & 898 \\
\hline
\end{tabular}
\end{table}

The quality knob is one such margin, and its payoff tracks the load.
Against a faithful GenServe-like baseline~\cite{genserve} (identical preemption and
placement, no quality knob), the knob is decisive under contention. At
the tight $600$\,ms rung the no-knob arm serves \emph{zero} bulk while
degrade-to-floor opens $0.79$\,blk/s/device, and the payoff decays
$1.96{\times}/1.46{\times}/1.21{\times}$ toward unity as slack appears.

\section{Related Work}
\label{sec:related}
 
\paragraph{Multi-SLO LLM serving.} Continuous batching and
SLO-differentiated scheduling (vLLM~\cite{vllm}, Orca~\cite{orca},
Sarathi-Serve~\cite{sarathi}, SGLang~\cite{sglang}, the 2026 multi-SLO
wave~\cite{flowprefill}) optimize token streams whose sessions are
stateless beyond KV prefixes, and they treat the cache as cheaply
reducible, evictable, quantizable, and prefix-shareable\ifextended{} (H2O~\cite{h2o} evicts it, CacheGen~\cite{cachegen} encodes it, KVQuant~\cite{kvquant} quantizes it, and vLLM and SGLang share it by exact prefix~\cite{vllm,sglang})\fi, precisely the handles the
kill-family (\S\ref{sec:measure}) shows this state denies.
These serving stacks differ from \sys{}: world-model serving inverts
the premises (fixed-shape
diffusion quanta, a multi-GB resident working set with fork semantics,
quality as a contracted knob), so schedulers tuned on token economics
inherit none of the placement, window-lifetime, or determinism
structure here.
\paragraph{LLM KV-cache live migration.} Most similar to \sys{} is
Llumnix~\cite{llumnix}, which achieves $20$--$30$\,ms sequence-length-independent
downtime as a property of the \emph{object}, since an append-only KV
cache copies its immutable prefix in parallel with decode; concurrent
systems migrate LLM KV state for reconfiguration~\cite{remp} and for
spot-cluster continuity~\cite{shuntserve} under the same append-only
assumption. A
window-rewrite world has no immutable prefix (\S\ref{sec:measure-dirty}) and tolerates neither
approximate copies nor destination recompute (\S\ref{sec:measure}), so
WorldMove is the primitive for the regime where both escapes fail.

\paragraph{Disaggregated LLM serving and commodity KV transport.}\looseness=-1{}
Disaggregated serving splits a request across prefill and decode pools and ships the KV cache between them\ifextended{} (Splitwise~\cite{splitwise}, DistServe~\cite{distserve})\fi. Production stacks
commoditize the byte transport itself: Mooncake~\cite{mooncake}
maintains a datacenter-wide KVCache pool with topology-aware RDMA
transfer, and NIXL~\cite{nixl} offers pluggable point-to-point
transfer. What they move is the same object: append-only and
lossy-tolerant. This paper's object forces the opposite, a
deadline-bounded, all-or-nothing motion of a heavily-mutating state
with no recompute escape. \sys{} adds no transport novelty to this stack: these movers are
candidate data planes beneath WorldMove, and the contribution is the
admissibility and verification layer that decides whether a move is
legal at all.
\paragraph{On-device execution-state capsules.} Concurrent and
independent work, Execution-State Capsules / FlashRT~\cite{flashrt},
snapshots and restores the complete live execution state of a
graph-bound model \emph{byte-exactly} for on-device serving. Its
key-value-only ablation independently echoes our kill-family
(\S\ref{sec:measure}), so we claim no novelty for byte-exact
whole-state capture in itself. These primitives resemble \sys{} up to
the boundary of \emph{movement and
fleet scheduling}: byte-exact whole-state capture exists on-device
(FlashRT), bit-exact cross-hardware re-execution exists offline via
software emulation~\cite{bitexactverif}, and cross-GPU numerical
reproducibility is an active target for LLM inference~\cite{heal}, but
none of these moves live state between devices under an interactive
deadline. That intersection,
deadline-bounded motion of an irreducible resident state under the
conditions it forces ($\mathrm{BW}\ge\mathrm{cache}/L$, $N\le HB/C$), is
this paper's subject.
\paragraph{Real-time scheduling.}\looseness=-1{} Our guard bands, degradation tiers,
and (m,k)-style miss accounting~\cite{mksched} descend from
mixed-criticality~\cite{vestal} and imprecise-computation
theory~\cite{imprecise}. We contribute the measured constants and
regime boundaries, including a GPU-quantized instance of slack
stealing~\cite{slackstealing} (a $94$\,ms \emph{non-preemptible}
quantum turns $24$\,ms of nominal slack into zero usable bulk) and a
critical section no priority-inheritance
protocol~\cite{priorityinheritance} can shorten, so the one
feasibility knob is the contractual quantum-length cap the guard
band enforces. D3~\cite{d3} and
PDQ~\cite{pdq} schedule deadline-flows that are preemptible, divisible,
and unverified. These deadline-flow schedulers differ from \sys{}:
atomic state motion removes all three relaxations and imports
the non-preemptive blocking term (\S\ref{sec:law-fleet}).
\paragraph{Diffusion/video serving\ifextended{} and game streaming\fi.} \ifextended DiT
serving~\cite{xdit} and GENSERVE~\cite{genserve} schedule stateless
requests, and StreamDiffusionV2~\cite{streamdiffusion2} makes one stream
fast where we price many sessions' co-residency.
\fi TurboServe~\cite{turboserve} and
SlackServe~\cite{slackserve} (both concurrent work)
establish that streaming video generation is a serving problem with
playout deadlines, and TurboServe makes live migration of \emph{stateful}
text-to-video sessions routine fleet control ($23$--$30$\,ms
RDMA/NIXL chunk-boundary moves). \ifextended Its when-to-migrate is a
greedy gain heuristic, and because its sessions are prompt-driven streams it
never characterizes how much of the moved state is rewritten. \fi TurboServe resembles \sys{} at the
mechanism level, but its protocol confirms only that the required buffers
were installed and its migration trigger carries no deadline-gated
admissibility. This paper supplies the
identity contract, the admissibility condition, and the object
characterization a fleet depending on moved state needs. \ifextended The 2026
world-model acceleration line (WorldKV~\cite{worldkv},
X-Cache, WorldCache~\cite{worldcache}) targets an
orthogonal problem to \sys{}: it operates
\emph{inside} one session's inference and leaves admission, placement,
and cross-tenant deadlines open, the layer this paper supplies.\fi
X-Cache~\cite{xcache}, built to approximate, still recomputes KV-update chunks to
keep error out of the autoregressive cache, the same
no-safe-approximation wall our kill-family measures from the outside
(\S\ref{sec:measure}).
\paragraph{State movers: liveness vs.\ contractual identity.} Treating
running state as movable is a deep tradition. \ifextended Live VM
migration~\cite{clark2005migration} hands off approximately over a
small writable working set, CRIU~\cite{criu} checkpoints exactly but
cold, Gandiva~\cite{gandiva} shares a GPU cluster and never
moves, and the replay/redo/SMR line recomputes state from cheap
causes and presupposes platform determinism. All optimize
\emph{downtime} under a liveness criterion their objects can afford.
\fi CRIUgpu~\cite{criugpu}, the nearest GPU instantiation, restores
from disk in $38.8$--$145.1$\,s with no bit-exactness contract (vs.\
$18.8$\,ms in-fabric motion under a bit-identity contract). The move itself is becoming
vendor infrastructure: NVIDIA's \texttt{cuda-checkpoint} ships GPU
migration for cluster scheduling~\cite{cudackpt} and serverless
platforms productize GPU snapshot/restore, yet none states an
integrity, determinism, or bit-exactness guarantee, and the restore
products pin restores to matching hardware, the
architecture-class boundary this paper measures and schedules around.
These movers differ from \sys{} on the identity axis: exact-state elasticity admits a move only when a \emph{bit-identical} resume is
reachable before the deadline. A world
model must \emph{manufacture} its determinism, and, at native rate,
only within an architecture (\S\ref{sec:eval}), so none of these
composes the primitive here (full lineage: TR~\S E).

\paragraph{Packing and finite-source loss.} \sys{}'s pool sizing resembles the classical models
(Engset~\cite{engset}, Erlang~\cite{erlang}, setup
costs~\cite{gandhi,phungduc}). The contract changes only which primitive
the model must include: bit-exact migration.

\section{Limitations and Open Problems}
\label{sec:limits}
\textbf{Scale.} All multi-GPU results are single-node ($\le$$8$ GPUs). The
fleet crossover $N_{\max}{>}1$ is measured at full size on a
$100$\,GbE fabric under serialized contention (\S\ref{sec:eval-verif});
cross-node \emph{placement} is modeled only in simulation.

\looseness=-1{}\textbf{Migration mechanism.} The atomicity unit is the block boundary.
Sub-block yield points, delta encodings for the append-only strip, and GPUDirect staging are engineering headroom.

\textbf{Demand realism.} The demand ladder is parameterized from recorded production
traces (Azure LLM Inference~\cite{azurellm}, BurstGPT~\cite{burstgpt}) and public
interactive-session statistics (\S\ref{sec:eval}), so it captures the pre-deployment
\emph{shape} of the problem. What is not public is a \emph{replayable} world-model
production trace\ifextended{} (the one production operator~\cite{turboserve} publishes
only per-minute aggregates)\fi, so a validated production mix does not yet exist.
\ifextended Table~\ref{tab:necessity} (Appendix~\ref{app:boundary}) records which
present workloads this applies to and why.\fi

\looseness=-1{}\textbf{Backend breadth.} The quantitative spine is one backend family
(Matrix-Game-2) plus a calibration row on a second (open-oasis);
abstraction boundaries are measured on $N{\le}2$ backends and labeled accordingly. The compress-vs-fidelity cliff, though, tracks the model's positive-Lyapunov regime and reproduces in two synthetic chaotic systems under the same codecs, which points to a mechanism beyond an MG2-only artifact.

\textbf{Secondary placement effects.} Residual core/IRQ contention
($12$--$35\%$ spread on correctly NUMA-bound devices) remains uncontrolled.
Per-engine core sets are future work.

\textbf{Quality horizon.} Both sustained tiers drift to unusable by
$\sim$$10$\,s of continuous generation on MG2 (self-measured; tier
detail, TR~\S E), a
model-layer constraint the fleet inherits\ifextended, surfaced as the scheduled
maintenance class (C5)\fi. Scheduling cannot fix it.

\textbf{Cross-arm comparability.} The causal comparison (Table~\ref{tab:abc})
has a modest trial count, and its blind baseline is non-degrading by
construction. A four-arm decomposition on a second platform
(8$\times$A100-80G, same operating point, three trials) separates the two
factors: blind round-robin in place of the contract scheduler costs
$3.4$\,pp of C1 miss with degradation enabled, while enabling degradation
under blind scheduling recovers only $1.2$\,pp, an ordering that holds in
every trial. The contrasts are policy bundles rather than a clean factorial:
feasibility arithmetic pins both contract arms at the $2$-step floor, the
blind arm's realized degrade dose is scheduler-dependent ($4.5\%$ of
blocks), and blind-arm misses concentrate in the initial bulk-drain
transient, so the magnitudes are specific to $10$-minute cells. In this
regime prioritization carries most of the advantage; the quality knob is
secondary.

\clearpage
\appendix
\section*{Appendices}

\section{Verification-plane forensics}
\label{app:forensics}

This appendix records the evidence behind the fan-in freeze
regime of \S\ref{sec:eval-verif}. The decision-grade constants and
Figure~\ref{fig:freeze} (raw per-chunk timestamps from the 2026-07-04
fabric) stay in the body.
\paragraph{The episode anatomy.} Solo, concurrent CRC already slows the wire $12\%$
and leaves a $\sim$$1/8$ stall tail, and $2\to1$ fan-in amplifies the
coupling into instability on this receive path. In $2/3$ of the barrier-synchronized incasts the wire enters discrete
second-scale freezes beginning exactly where the first CRC burst overlaps
an active receive (per-chunk timestamps, both collapsed incasts; flows stall synchronously and recover in-transfer),
and the two controls isolate the CRC work as the differing variable: a no-op executor arm with
the identical dispatch shape is $3/3$ stable at $267$\,ms wire, and
sequential verification is $3/3$ stable ($405$--$409$\,ms sender e2e).
Capping the CRC engine to quarter width does not remove the episodes
($2/4$ repeats trip at $8$ threads), ruling out aggregate memory bandwidth.
Under fan-in the CRC load backpressures the NIC into receive-path loss.

\paragraph{The counter inference chain.} Receive-queue discards fire on
every episode, a rendezvous-read packet is dropped (inferred from the
flanking counters), and the RC transport waits out its acknowledgment
timer. The flows ran as configured on default CloudLab RoCE, on the
default lossy priority: the discards land there, and zero ECN-marked
RoCE packets were recorded.

\paragraph{The timeout-setting record.} The freeze constant
measures $2147$\,ms $=2\times(4.096\,\mu\mathrm{s}\cdot2^{18})$ at the
$1$\,s default, and quartering the configured timeout moves it to
$536$\,ms $=2\times(4.096\,\mu\mathrm{s}\cdot2^{16})$, a
$4.006\times$ shift, with the requester's
\texttt{local\_ack\_timeout\_err} counter firing on $18/18$ stalled
incasts. A second, continuous stall component ($1.2$--$2.1$\,s, capped at
the constant) rides with it. Episode \emph{probability} is
fabric-state-dependent ($2/3$ of incasts in one hour, $12/12$ in another),
and the knob shrinks the constant $4.006\times$ (whole episodes
$\sim$$3.7\times$) but prevents none. Further quartering does not: the NIC (the d6515's dual-port ConnectX-5) enforces a minimum timeout exponent ($2^{16}$), so on this fabric an episode always overruns the $400$\,ms block regardless of tuning. Configured
settings from $60$ down to $4$\,ms all clamp at the same
$536$\,ms. The clamp is what
isolates the NIC's minimum timeout exponent.

\paragraph{Invisible from the IP layer.} The state is invisible from above: across $4.3$ hours of concurrent probing, the fabric's latency plane stayed clean ($14{,}973$ ICMP samples, max RTT $0.441$\,ms, zero loss, episode and clean windows statistically indistinguishable) while the bulk plane sat in multi-second freezes. Only the bulk transfer with byte-identity checking exposes the regime.

\paragraph{The capacity arithmetic (\S\ref{sec:eval-verif}).} The
crossing charges the receive host's checksum capacity against aggregate
ingest: $B_{\mathrm{crc}}{=}52.6$\,GB/s is the $32$-thread engine of
Appendix~\ref{app:swfloor}, $B_{\mathrm{wire}}{=}12.5$\,GB/s is one
$100$\,Gb wire, and the crossing sits at
$B_{\mathrm{crc}}/B_{\mathrm{wire}}{=}4.2$ wires ($1.05$ at
$400$\,Gb/s). The episodes recorded here fire with aggregate ingest
fair-shared to a single wire, $\sim$$24\%$ of that capacity, which is
what places the measured failure on the timer boundary rather than the
capacity boundary.

\section{Weaker contract tiers}
\label{app:tiers}

Exact-state motion is forced only under the exact (T0) contract on a
readout-irreducible backend. The weaker T1/T2 tiers (taxonomy: TR~\S E)
are the regimes where the state the contract needs is smaller than $|S|$
given what the destination holds, so lossy compression or replay become legal
primitives. The tiers are measured where they surface in the body: the
cross-architecture readout of \S\ref{sec:eval-m5} lands at a bounded T1,
and weaker contracts restore a reducible subset (\S\ref{sec:eval}).

\section{Fleet response-time test development}
\label{app:fleettest}
The specialization behind the fleet test of \S\ref{sec:law}, in full.
A fleet migrates many sessions over a \emph{shared} channel. Each move
runs to commit or abort within its horizon, because yielded time is
unrecoverable under the deadline gate, so concurrent migrations block and
interfere as non-preemptive real-time jobs do and the condition lifts to
the standard deadline-monotonic response-time test~\cite{jeffay91}
specialized to atomic state motion: session $i$ moves $C_i=S_i+D_iH_i$
bytes (state bulk plus horizon dirtying) and is schedulable when
$R_i = C_i/B + \mathrm{blocking}_i + \mathrm{interference}_i \le H_i$,
with the usual one-longer-transfer blocking and shorter-deadline
interference terms.

\section{Serving-cell detail}
\label{app:serving}

The three serving-cell economies summarized in \S\ref{sec:measure}, in
full.
\emph{(i) Quality is a priced ladder with bifurcated semantics:} denoise
steps price linearly (a 2-step block costs $3/4$ of a 3-step one), and what
degradation \emph{means} splits by contract: self-coherence contracts
absorb it (drift slope unchanged within the pre-registered $15\%$
line, $0.0147$ vs $0.0139$), while
reference-tracking contracts compound (tail-to-first-window LPIPS $2.78$) and must cap
consecutive degraded blocks, $(m,k)$-firm style.
\emph{(ii) Decode is a second workload:} full-VAE decode misses every
$30$\,fps deadline in every arm (the budget is infeasible regardless of
policy) and a distilled tiny decoder ($1.1$--$4.3$ vs
$20.8$\,ms/frame) is what opens $30$\,fps at all (tiering: TR~\S E).
\emph{(iii) State is structured, and the structure is policy-relevant.}
Truncating the spatial KV's temporal window to half is visually graceful
where zeroing it is not, and action caches are negligible, so eviction can act at component granularity. But truncation discards \emph{old} context; approximating the
\emph{current} resident set has no safe direction at the compressor's error
scale: perturbations of the median codec-error norm along unstable,
stable, and readout-orthogonal directions alike collapse fidelity by
$\ge$$60$\,dB (means $73$--$92$\,dB, all $12$ profiled blocks, vs a
$2$\,dB tolerance). The working set is quality-compressible yet not
\emph{lossily} reducible under the divergence contract. Co-batching, the LLM staple, is
a weak knob here: B$=$4 buys 1.16$\times$ throughput at 3.5$\times$ the
quantum. Finally, feasibility rows are \emph{operating points} rather than GPUs
(one 5090 clock-pinned 2.8$\to$1.2\,GHz traverses $2.2\times$ in quantum),
and marginal cells (slack $1.0$--$1.15$) do not survive contention.

\section{Software-floor detail}
\label{app:swfloor}

The forensic decomposition behind the first-generation mover's
$251$\,ms / $53$\,Gbps datum of \S\ref{sec:eval-verif}.

\begin{figure}[t]
  \centering
  \includegraphics[width=\columnwidth]{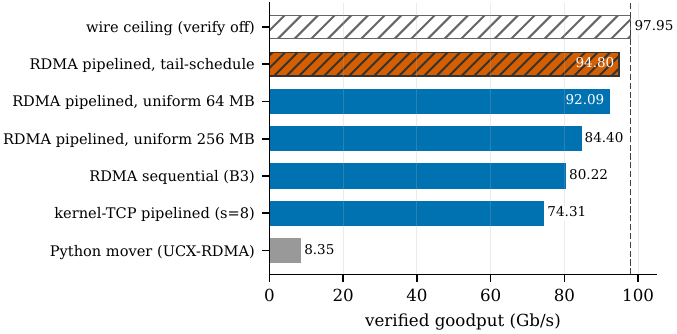}
  \caption{Verified-transfer price list, measured on one fabric on one
  day (two CloudLab d6515s, $100$\,GbE, 2026-07-05): with off-path
  pipelined verification on by default, verified throughput reaches
  $92.09$\,Gb/s at uniform $64$\,MB chunks ($94\%$ of the $97.95$\,Gb/s
  wire ceiling) and $94.8$\,Gb/s with the tail schedule.}
  \label{fig:placement}
\end{figure}

\paragraph{Registration and CRC-engine decomposition.} With the
session's slabs registered once (a first transfer pays an
amortizable $\sim$$1.8$\,s registration charge) the Python transport holds $97\%$ of
verbs. Within the $251$\,ms move,
verification costs $110$\,ms ($44\%$; a $32$-thread CRC engine
sustains $52.6$\,GB/s, floor $\sim$$30$\,ms).

\paragraph{The RoCE tier and the shape below it.} On a
$25$\,Gb RoCE tier the same object rides RDMA at $21.5$\,Gbps steady
\emph{including} the byte-identity check and fail-closed ack, $93\%$ of
the fabric's verbs ceiling ($23.15$\,Gbps), verification costing $7\%$
of wire time. The two measured residuals are a one-time registration
charge and a $3.4\times$ no-peer-DMA staging penalty.
(A six-point \texttt{netem} sweep, $0.5$--$10$\,Gbps: every arrival
bit-exact, affine with slope $\approx$$1$, so the whole object ships, no
clean dirty set, \S\ref{sec:measure-dirty}; the $3.2$\,s intercept is
our Python-socket serialization floor rather than the object's, which is
$18.8$\,ms same-node, \S\ref{sec:eval-m4}.)

\section{Applicability boundary}
\label{app:boundary}

The applicability audit of \S\ref{sec:limits}, in full.

\begin{table}[t]\centering\footnotesize
\caption{Applicability boundary (our own audit): the escape that lets each
present workload avoid exact-state fleet elasticity, and the measured or
structural force that closes it. The paper targets the emerging
persistent-frame-model regime, where the escapes are already closing (full
grid in the artifact).}
\label{tab:necessity}
\setlength{\tabcolsep}{3pt}\footnotesize
\begin{tabular}{@{}L{2.05cm}L{2.35cm}L{2.7cm}@{}}
\toprule
workload & escape today & what closes it \\
\midrule
deterministic sim & replay cheap \& exact & neural sims inherit the \\
\ (CARLA class) & & divergence we measure (\S\ref{sec:eval}) \\
RL / rollout farms & episodic resets, & persistent worlds: replay \\
 & externalize+replay & cost grows $O(T)$ \\
offline data factories & batch; re-run the job & interactivity (deadlines) \\
consumer interactive & state decoheres within & the coherence race (unbounded- \\
\ (Oasis/Genie class) & minutes; evict \& restart & persistence claims now shipping) \\
streaming T2V serving & state tolerates unverified & exactness-bearing tenants: replay, \\
\ (production~\cite{turboserve}) & moves; no contract layer & forks, eval; window-rewrite worlds \\
\bottomrule
\end{tabular}
\end{table}

\section{Cross-architecture readout detail}
\label{app:crossarch}

The measurement behind the bounded-T1 cross-architecture verdict of
\S\ref{sec:eval-m5}: two
boundaries, A100$\times$H100 and A100$\times$L40S, the software stack held
identical down to the cuDNN patch, three seeds per pair. Across the $174$ post-transient
blocks the readout holds a median PSNR of $61$\,dB (minimum $43$\,dB, never
below a $30$\,dB serving floor, growth exponent
$\lambda\in[-0.001,+0.023]$). (Block~0 is a $6$--$14$\,dB initialization
transient, the sole excluded block, reported not hidden.) Two pairs sharing
one A100 reference are a small taxonomy; the claim does not extend to all accelerators.

\section{Churn and consolidation detail}
\label{app:churn}
\paragraph{Live-churn convergence (\S\ref{sec:eval-m4}), in detail.} Two
H100s run the shipped iterative pre-copy API against a generating
session. Round~0 ships all $399$
four-MB chunks, and every subsequent round's dirty set saturates at $310$
chunks ($0.777$ of the cache, which reproduces the independently measured
$\rho_{\mathrm{WWS}}$ in real time), so pre-copy cannot converge
while the model generates and converges the moment generation pauses at a
block boundary ($1.14$\,s downtime including the frozen residual, dual CRC,
and fail-closed install). A
second engine adopts the migrated cache and its next three generated blocks
are bitwise identical to the source's counterfactual continuation, block for
block.

\paragraph{Certification detail (\S\ref{sec:eval-m4}).} On the
2$\times$A800 cell the pre-warmed install lands with $0\%$ added
deadline misses vs.\ $0\%$ for the non-migrated control at the
$600$\,ms contract. Two oracles confirm continuity: the cache content hash is bit-identical across the move, and the migrated stream's output matches a natively-run session block for block. A negative control (perturbing the cache before resume) diverges on every subsequent block.
Silent corruption at fleet scale is
documented hardware reality~\cite{dixit,hochschild}, and it occurred in
these runs: one rented host's direct-P2P path silently corrupted bytes
mid-transfer (IOMMU/ACS misconfiguration, a failure mode NVIDIA
documents for this configuration family: the IOMMU ``must be
disabled\ldots to prevent silent device memory corruption'' on
PCIe P2P~\cite{cudaguide}). The CRC check rejected the transfer before
any state was served (artifact: \texttt{mig\_realstate}).

\paragraph{The coupon-law derivation (\S\ref{sec:eval-m4}).} On the $100$\,Gb
fabric the dirty set follows the
occupancy/coupon-collector expectation, $\mathrm{distinct}=N(1-e^{-w/NC})$, within $4\%$ over $30$+ round
observations. Its calibrated recursion \emph{pre-registered} the
boundary. The recursion predicted convergence in $11/20$ rounds at
$3.0/3.5$\,GB/s and divergence above $D^{*}{=}3.85$\,GB/s; the runs
measured $12/18$ rounds and a cap-$40$ divergence at $4.2$, $3/3$.

\paragraph{Swap-tier detail (\S\ref{sec:eval-fleet}).} The $11$\,GB/s host
tier is a conservative modeling
band (measured anchors run $12.5$--$56$\,GB/s). Swap adds nothing below
$Z{\approx}26$\,s, and a simulated NVMe tier beats residency only past
$Z{\approx}340$\,s at $7$\,GB/s. Parking density is priced on the movable
set, which alone is sufficient for bit-exact resume on a warm engine
(\S\ref{sec:eval-m4}); the $5.32$\,GB \emph{active} footprint of
\S\ref{sec:measure} bounds concurrent generation rather than parking
density.

\paragraph{PCIe-5 atoms and interference arms (\S\ref{sec:eval-fleet}).}
Restore total $34.7$\,ms (medians; resume wire $29.60$ $+$
rehydrate $2.79$ $+$ install $2.31$) vs.\ the modeled $154$\,ms, with
suspend at $29.78$\,ms, on a $32$\,GB part with an $11$-session residency
floor. The $K^{*}$ knee location is DES-confirmed within $0.2\%$, and
the CTMC itself is DES cross-checked at $1.5$--$2.8\%$ at three points.
Background arrival-CRC inflates foreground generation
$+13.0\%$ when one PCIe link carries both planes (host-only CRC
$+0.67\%$, below the run's pre-registered $1\%$ noise floor; concurrent suspend/resume
$+75.5\%$, so the diagram's copy-does-not-contend simplification fails on
single-link hosts; separate links remove the shared substrate). Every
interference arm ended with byte-identical latent CRCs.

\section{Envelope detail}
\label{app:envelope}

\begin{figure}[t]
  \centering
  \includegraphics[width=0.88\columnwidth]{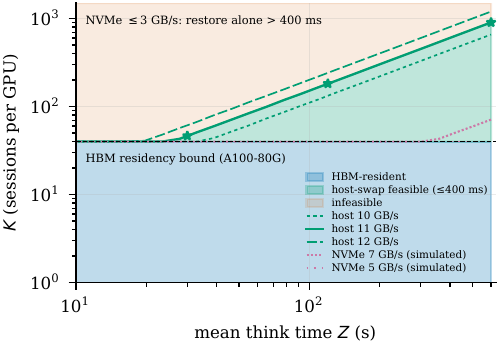}
  \caption{Consolidation phase diagram: sessions per device $K^{*}$
  under the $400$\,ms/$1\%$ SLO versus think time $Z$ (machine-repairman
  CTMC with measured cost constants; sessions park at $1.67$\,GB cores).}
  \label{fig:kmax}
\end{figure}

\begin{figure}[t]
  \centering
  \includegraphics[width=0.9\columnwidth]{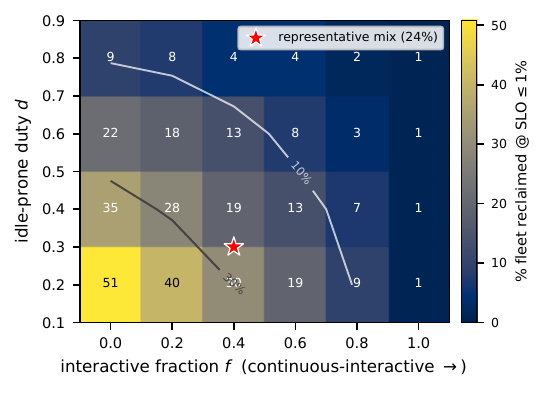}
  \caption{Fleet reclaimed at a $1\%$ reactivation SLO bound over
  interactive fraction $f$ and idle-prone duty $d$ ($C_m{=}96.4$\,ms):
  up to $51\%$ in the idle-prone corner, $\sim$$0\%$ at
  continuous-interactive load.}
  \label{fig:migphase}
\end{figure}

\paragraph{Reclaimable population and thresholds
(\S\ref{sec:eval-fleet}).} The reclaimable population is suspended
sessions: parked forks, between-round evaluation, paused bulk. The
empirical-duty sizing is an exact Poisson-binomial convolution
($N{=}114$, $1\%$ reactivation SLO) over the gaming rung's per-session
duty distribution. The policy curve spans $6.1$--$19.3\%$ as the suspend
threshold sweeps $30$\,s down to $2$\,s, the majority-interactive cell
uses idle-prone duty $0.35$, and the reclaimable population is the AFK
tail, the near-idle bottom decile the synthetic assumption erases. The
warm buffer is an Erlang buffer of $\sim$$2$ device equivalents at $120$
sessions. Aggressive thresholds ($5$\,s) generate a reactivation flow
two orders of magnitude larger and are viable only where suspend
returns a warm engine slot rather than releasing the device; that path
is mechanically cheap: suspending the full cache to pinned host memory
and resuming it costs $29.5$\,ms each way on a PCIe-5 workstation,
bit-exact through the cycle, $\sim$$1.2\%$ overhead against a $5$\,s gap.

\paragraph{Two-instance move paths (\S\ref{sec:eval-xnode}).} Both
moves ship the movable set \emph{from a live GPU cache}
(device$\to$host$\to$multi-stream) under a single global CRC. Raw
records are in the artifact.

\paragraph{Admission-comparison detail (\S\ref{sec:eval-xnode}).} The
idiom-stale arm under-modeled because at $\rho\approx1$ queueing
compounds service-time error. A $7\%$ overestimate was harmless while a
$30\%$ underestimate collapsed the arm, placing the tolerance cliff
somewhere between. The Jacobson-style estimator's transplant is not
free: TCP's single-sample variance initialization rejects everything
under a $400$\,ms deadline, admission needs its own slow-start, and the
estimator trades throughput for margin at overload. In the two-fabric
replication, a second pair reproduces the controller row to the integer
at every rate and the cold-idiom calibration ratios match within
$1\%$, so the placement inversion is a property of the deployment
idiom rather than of one machine.

\section{Artifact statement}
\label{app:artifact}

The artifact ships claims files with falsifiability
pointers, the exploration tree including dead ends, per-run provenance, and
a numbers manifest mapping every figure to its reducer. It accompanies
publication.

\end{document}